\documentclass{aa}
\usepackage{graphicx,natbib}
\def\bp{$\beta$\,Picto\-ris}

\def\hda{HD\,100546}

\def\psfhda{HD\,106797}
\def\mwc{MWC\,480}
\def\psfmwc{HD\,29646}
\def\hdb{SAO\,206462}
\def\psfhdb{SAO\,206463}
\def\cs{cir\-cum\-stel\-lar}
\def\co{\mbox{coronagraphic}}

\def\haebe{HAeBe}

\def\h{\hfill\break}

\newcommand{\uma}[1]{^{\mathrm{#1}}}
\newcommand{\dma}[1]{_{\mathrm{#1}}}
\begin{document}
\thesaurus{08 (19.42.1 \hda, \hdb, \mwc ; 19.25.1; 04.01.1; 05.01.1; 19.94.1)}
           %
           %
\authorrunning{J.C. Augereau et al.} 
\titlerunning{HST/NICMOS2 \co\ observations of \hda, \hdb\ and \mwc}
\title{HST/NICMOS2 \co\ observations of the circumstellar environment
  of three old PMS stars: \hda, \hdb\ and \mwc.  \thanks{Based on
    observations with the NASA/ESA Hubble Space Telescope, obtained at
    the Space Telescope Science Institute, which is operated by the
    Association of Universities for Research in Astronomy, Inc. under
    NASA contract No. NAS5-26555.}}  \author{J.C. Augereau\inst{1}
  \and A.M. Lagrange\inst{1} \and D. Mouillet\inst{1} \and F.
  M\'enard\inst{1,2}}
\offprints{J.C. Augereau}
\mail{augereau@obs.ujf-grenoble.fr}
\institute{ Laboratoire
  d'Astrophysique de l'Observatoire de Grenoble, Universit\'e J.
  Fourier / CNRS, B.P. 53, F-38041 Grenoble Cedex 9, France \and
  Canada-France-Hawaii Telescope Corporation, PO Box 1597, Kamuela, HI
  96743, USA} \date{Received {}; accepted {}} \maketitle
\begin{abstract}
  The close environment of four old Pre-Main Sequence stars has been
  observed thanks to the coronagraphic mode of the HST/NICMOS2 camera
  at $\lambda=1.6\,\mu$m. In the course of this program, the detection
  of a \cs\ annulus around HD\,141569 has already been presented in
  \citet{aug99b}.
  
  In this paper, we report the detection of an elliptical structure
  around the Herbig Be star \hda\ extending from the very close edge
  of the \co\ mask ($\sim$50\,AU) to 350--380\,AU (3.5--3.8$\arcsec$)
  from the star. The axis ratio gives a disk inclination of
  $51\degr\pm3\degr$ to the line-of-sight and a position angle of
  $161\degr\pm5\degr$, measured east of north. At 50\,AU, the disk has
  a surface brightness between 10.5 and 11\,mag/arcsec$^2$, then
  follows a $-2.92\pm0.04$ radial power law up to 250--270\,AU and
  finally falls as $r^{-5.5\pm0.2}$. The inferred optical thickness
  suggests that the disk is at least marginally optically thick inside
  80\,AU and optically thin further out. Combined with anisotropic
  scattering properties, this could explain the shape of a brightness
  asymmetry observed along the minor axis of the disk. This asymmetry
  needs to be confirmed.
  
  The \cs\ disks around \hdb\ and \mwc\ are not resolved, leading to
  constraints on the dust distribution. A tight binary system
  separated by only $0.32\arcsec\pm0.04\arcsec$ is nevertheless
  detected in the close vicinity of \hdb.  \keywords{Stars: \cs\ 
    matter -- Stars: \hda, \hdb, \mwc}
\end{abstract}
%
%
\section{Introduction}
Dust is present around a large fraction of Main Sequence (MS) stars,
as shown by IRAS \citep{aum84}. Moreover, spectroscopic
\citep[e.g.][]{que00} and photometric \citep{cha00} detections
indicate that planetary systems are quite common. In some cases, the
dust lifetime is so short compared to the star's age, that it has to
be replenished, probably through collisions between planetesimals or
through evaporation \citep{bac93}. The best example so far among these
second generation disks is the one surrounding \bp\ aged more than
20\,Myr \citep{bar99}. Recently, submillimeter images resolved the
dust emission around a few isolated MS stars with different spectral
types and older than 0.1\,Gyr \citep{hol98,gre98}.

Circumstellar disks are also present around young objects, as clearly
demonstrated by HST \citep[][for instance]{pad99}, and by ground-based
observations \citep{dut96}.  The latter detected \cs\ disks around a
sample of CTTs in the Taurus cloud, but not yet around post T Tauri
stars. This is in general agreement with \citet{zuc93} and more
recently \citet{hab99} who have shown the rapid decrease of the \cs\ 
dust mass as the stars evolve to the Main Sequence \citep[see
also][]{hol98}.  Gas infall is also observed towards a few Herbig
Ae/Be (hereafter \haebe) stars \citep[e.g.][]{gra96,dew99},
tentatively interpreted as the result of comet evaporation. In this
scheme, this could indicate that planetesimal formation in the nebula
is quite common and rapid, as theoretically expected.

Very few disks have been detected around \haebe\ and more evolved Pre
Main-Sequence (PMS) stars \citep[HD\,163296 and
\mwc,][]{man97a,man97b}. A recent study of HR\,4796, an 8\,Myr star,
has evidenced that the disk is at least partly second generation and
that meter-size bodies should already be present \citep{aug99a}.
Similar conclusions could apply for the HD\,141569 disk
\citep{aug99b}. In both cases, the analysis relies on detailed studies
of the resolved images and spectral energy distributions (SEDs).  From
a general point view, the nature of transient disks, the role of
replenishing planetesimals if any, and the most relevant physical
processes in these environments are still to be clarified.
 
The general evolutionary scenario, to be tested and precised is then
that~:\h 1) the protoplanetary disks around young, embedded stars
clear out before the star reaches the main sequence through dynamical
activity and interaction with stellar radiation and that\h 2) new
optically thin \cs\ disks may be sustained around older MS stars by
destruction processes among planetesimals (collisions, evaporation),
formed during the Pre-Main Sequence (PMS) phase.\h The transition
between those two phases, when the original young disk is eroded, is
critical to validate the above scenario.  Such a study can be
performed on disks around evolved PMS stars.  Critical points for this
study are the analysis of opacity to stellar radiation and collisional
processes time-scales. Both of them require detailed information on
the spatial distribution and amount of \cs\ dust.

During HST/Cycle 7, we probe in coronagraphic mode the environment of
four evolved PMS stars with the NICMOS2 camera.  As part of this
program, we reported the positive detection of an extended ($>400$\,AU
in radius) structure around HD\,141569 in \citet{aug99b}. The
detection was confirmed by \citet{wei99}.  We address in the present
paper our results for three other sources of large interest among old
PMS stars~: \hda, \hdb\ and \mwc.  After a brief description of the
observing strategy, journal of observation and reduction procedure
(section \ref{obs}), we present in section \ref{hd100} the results for
\hda. An extended structure is detected and interpreted in terms of
\cs\ dust arranged within an inclined disk. Detection limits are
derived from unresolved structures around \hdb\ and \mwc\ (sections
\ref{hd135} and \ref{mwc}) and in each case we discuss the
implications on the dust distribution. We also detail the close
vicinity of \hdb\ which exhibits a tight binary system.
%
\section{HST/NICMOS2 observations and data analysis}
\label{obs}
\begin{table*}[!tbph] 
\begin{center}
\caption{\label{log}
  Observing log. Note that \hdb\ and \psfhdb\ form a visual binary
  system (HD\,135344). References~:
  {\scriptsize$^{(a)}$}~\citet{wae90a,wae90b},
  {\scriptsize$^{(b)}$}~\citet{cou95}, {\scriptsize$^{(c)}$}~Simbad
  (CDS), {\scriptsize$^{(d)}$}~\citet{mye98}.}
\begin{tabular}[h]{rccccccc}
 & Star Name & Spectral Type & V mag & Date & Integration Time & Same Orbit ? \\
\hline
Object : & \hda\ & B9Vne & 6.68 & Nov 09, 1998 & 16$\times$47.958\,s $=$ 12\,m\,47\,s & \\
Reference : & \psfhda\ & A0V & 6.07 & Nov 09, 1998 & 16$\times$47.958\,s $=$ 12\,m\,47\,s & Yes \\
\hline
Object : & \hdb\ & F4Ve{\scriptsize $^{(a)}$} ; F8V{\scriptsize $^{(b)}$} & 8.65 & Aug 22, 1998 & 7$\times$111.931\,s $=$ 13\,m\,03\,s & \\
Reference : & \psfhdb\ & F2{\scriptsize $^{(c)}$} ; A2{\scriptsize $^{(d)}$} & 7.9 & Aug 22, 1998 & 7$\times$191.961\,s $=$ 22\,m\,24\,s & Yes \\
\hline
Object : & \mwc\ & A2 & 7.72 & Feb 24, 1998 & 11$\times$39.953\,s $=$ 7\,m\,20\,s & \\
Reference : & \psfmwc\ & A2V & 5.73 & Feb 24, 1998 & 11$\times$31.959\,s $=$ 5\,m\,52\,s & Yes \\
\hline
\end{tabular}
\end{center}
\end{table*}
\begin{table*}[!tbph] 
\begin{center}
\caption{\label{paraastro}
  Some known astrophysical parameters of the target sources presented
  in this paper. References~: {\scriptsize $^{(a)}$}~\citet{van97},
  {\scriptsize $^{(b)}$}~\citet{syl96},
  {\scriptsize $^{(c)}$}~\citet{zuc95}, {\scriptsize
    $^{(d)}$}~\citet{cou95}, {\scriptsize $^{(e)}$}~\citet{dun97},
  {\scriptsize $^{(f)}$}~\citet{van98}, {\scriptsize
    $^{(g)}$}~\citet{man97b}.}
\begin{tabular}[h]{ccccccc}
\hline
Star Name & Distance {\scriptsize [pc]} & Luminosity {\scriptsize [L$_{\odot}$]}& Mass {\scriptsize [M$_{\odot}$]} & $T\dma{eff}$ {\scriptsize [K]} & Age [Myr] & Other Name\\
\hline
\vspace{-0.05cm} & \vspace{-0.05cm} & \vspace{-0.05cm} & \vspace{-0.05cm} & \vspace{-0.05cm} & \vspace{-0.05cm} & \vspace{-0.05cm} \\
\hda & $103^{+7}_{-6}${\scriptsize $^{(a)}$} & $32.4^{+4.5}_{-3.5}${\scriptsize $^{(a)}$} & $2.4\pm 0.1${\scriptsize $^{(a)}$} & $10470${\scriptsize $^{(a)}$} & $> 10${\scriptsize $^{(a)}$} & SAO\,251457 \\
\hdb & $84${\scriptsize $^{(b)}$} ;  $100${\scriptsize $^{(c)}$} &  &  & $6250${\scriptsize $^{(d)}$} ; $6660${\scriptsize $^{(e)}$} & & HD\,135344 \\
\mwc & $131^{+24}_{-18}${\scriptsize $^{(f)}$} & $32.4^{+13.3}_{-8.4}${\scriptsize $^{(f)}$} & $2.2\pm 0.3${\scriptsize $^{(f)}$} ; $2.3^{+0.1}_{-0.3}$ {\scriptsize $^{(g)}$} & $8710${\scriptsize $^{(f)}$} & $2.5^{+1.5}_{-0.9}${\scriptsize $^{(f)}$} ; $6${\scriptsize $^{(g)}$} & HD\,31648 \\
\hline
\end{tabular}
\end{center}
\end{table*}
\subsection{Target selection}
Our aim was to image \cs\ environments of transition (old PMS) objects
to establish the missing link between disks around young embedded
objects and $\beta\,$Pic-like stars in order to compare the disks with
those (optically thick) of young stars on the one hand, and that
(optically thin) of \bp\ on the other hand. Our four targets~: \hda,
\hdb, \mwc\ and HD\,141569 were selected according to the following
criteria~: \\
- to be PMS stars close to the MS, which can be characterized by a
large IRAS excess, together with an age estimation
(via photometry or spectroscopy) of about 10$^6$--10$^7$ years. \\
- to be surrounded by \cs\ dust (IR excess) and CO. CO detection
indeed provides a good indication for the gas to be extended.\\
- to have been observed at various wavelengths to allow comprehensive
investigation.
%
\subsection{Observing strategy}
The high level of contrast between the investigated \cs\ environments
and their central star implies the use of coronagraphic techniques.
The subtraction of residual Point Spread Fonction (PSF) wings is
required to take full benefit of the coronagraph's high contrast
capabilities. For that, we observed in the same optical configuration
a comparison star within a few degrees of the science target, of
similar or slightly higher brightness, with similar spectral type and
thought to be free of circumstellar matter.

Comparison stars need to be observed {\it close in time} to each
corresponding science target observation because of the slight
possible variation of the PSF with time. This method has proved to be
efficient for ground-based coronagraphic data with adaptive optics,
especially because it is still efficient if the PSF is not perfectly
symmetrical.  It also allows the detection of circumstellar disks far
from edge-on orientation.
\subsection{NICMOS2 data}
We obtained coronagraphic images of our four targets with the
HST/NICMOS2 camera between February and November, 1998. All the
observations were performed in Filter F160W ($\lambda_{c}=1.6\,\mu$m,
$\Delta\lambda=0.4\,\mu$m). For the three targets presented in this
paper, the observing log is summarized in Table \ref{log}. The
individual integration times were determined so as to stay below the
saturation limit of the detector on the brightest pixels (near the
central hole edge area).  Filter F160W was selected for optimum
detectability: the PSF is well sampled and the background (zodiacal
light and thermal background) remains low.
\subsection{Reduction procedure, photometry and astrometry}
\label{reduction}
For each observed star, the calibrated files (in counts/s) provided by
the STScI are coadded to form a single image of the star during a same
orbit. The reduction procedure then consists in subtracting the
comparison star carefully scaled to the star of interest. The
determination of the scaling factor is critical since a small change
can significantly modify the photometry or in the worst case induce
artifacts.  It is assessed by azimuthally averaging the profile
resulting from the division between the star of interest by the
reference star images \citep[see also][]{aug99b}.

We use the 2.077$\times$10$^{-6}$\,Jy\,.\,s\,/\,counts factor to
convert NICMOS count rates to absolute fluxes and a zero point flux
density of 1040.7\,Jy to convert to magnitudes (NICMOS data handbook,
version 4.0, dated december 1999). Point source photometry is obtained
by integrating the total flux within a 0.5$\arcsec$ radius circular
aperture, then applying a 1.15 correcting factor to compensate for the
flux which fell out of this aperture (NICMOS photometry update web
page).

The pixel scale slowly varies with time but stays within the range
0.075$\arcsec$--0.076$\arcsec$ in average.
%
\section{\hda }
\label{hd100}
\subsection{A Herbig Be star close to the ZAMS}
The \hda\ star has been intensively studied since it was identified as
a member of the \haebe\ group \citep{hu89}. Indeed, \hda\ is a B9Vne
star showing a strong infrared excess peaked at about 25\,$\mu$m
(IRAS) due to \cs\ material and then fulfills the criterions proposed
by \citet{wat98} to identify \haebe\ stars. Some astrophysical
parameters are summarized in Table \ref{paraastro}.

The position of the star in the HR diagram indicates that \hda\ is
close to the ZAMS leading to an estimated age larger than 10\,Myr
\citep{van97}. This star, associated with the dark cloud DC\,296.2-7.9
\citep{hu89,vie99}, is surrounded by a large amount of dust: between a
few tens \citep{bou00} and a few hundred Earth masses \citep{hen98}.
The dust is rich in C and O and in particular PAH and silicates
\citep{mal98}.  The latter are very valuable since they constrain the
dust optical properties. ISO observations also evidenced similarities
between the 10\,$\mu$m emission of the \cs\ dust and that of comet
Hale-Bopp \citep{cro97,mal98}.

As commonly observed for other \haebe\ stars, \hda\ exhibits
photometric, polarization and spectroscopic variability
\citep{gra96,gra97,van98,yud98,cla99,vie99}. The interpretation of
such events in particular in terms of star-grazing comets
\citep{gra97,vie99} remains however uncertain \citep{lag00,beu00}.

Another issue concerns the very distribution of the material
surrounding \hda. The disk detection in scattered light between 40
(0.4$\arcsec$) and 200\,AU (2$\arcsec$) by \citet{pan00} reveals a
$\sim$ 50$\degr$ inclined disk almost in the SE-NW direction.  At
$\lambda=1.3$\,mm, an extension as large as 15--20$\arcsec$ in radius
and close to detection limit is also reported by \citet{hen98} in the
same direction. From the theoretical point of view, it is unclear
whether the \cs\ material arranges within a single disk.  For
instance, the presence of an additional envelope is proposed to
reproduce the full spectral energy distribution \citep{hen94} and to
explain specific spectroscopic events \citep{vie99}.
\subsection{Results}
\begin{figure*}[!tbp]
\begin{center}
\includegraphics[angle=-90,origin=bl,width=0.95\textwidth]{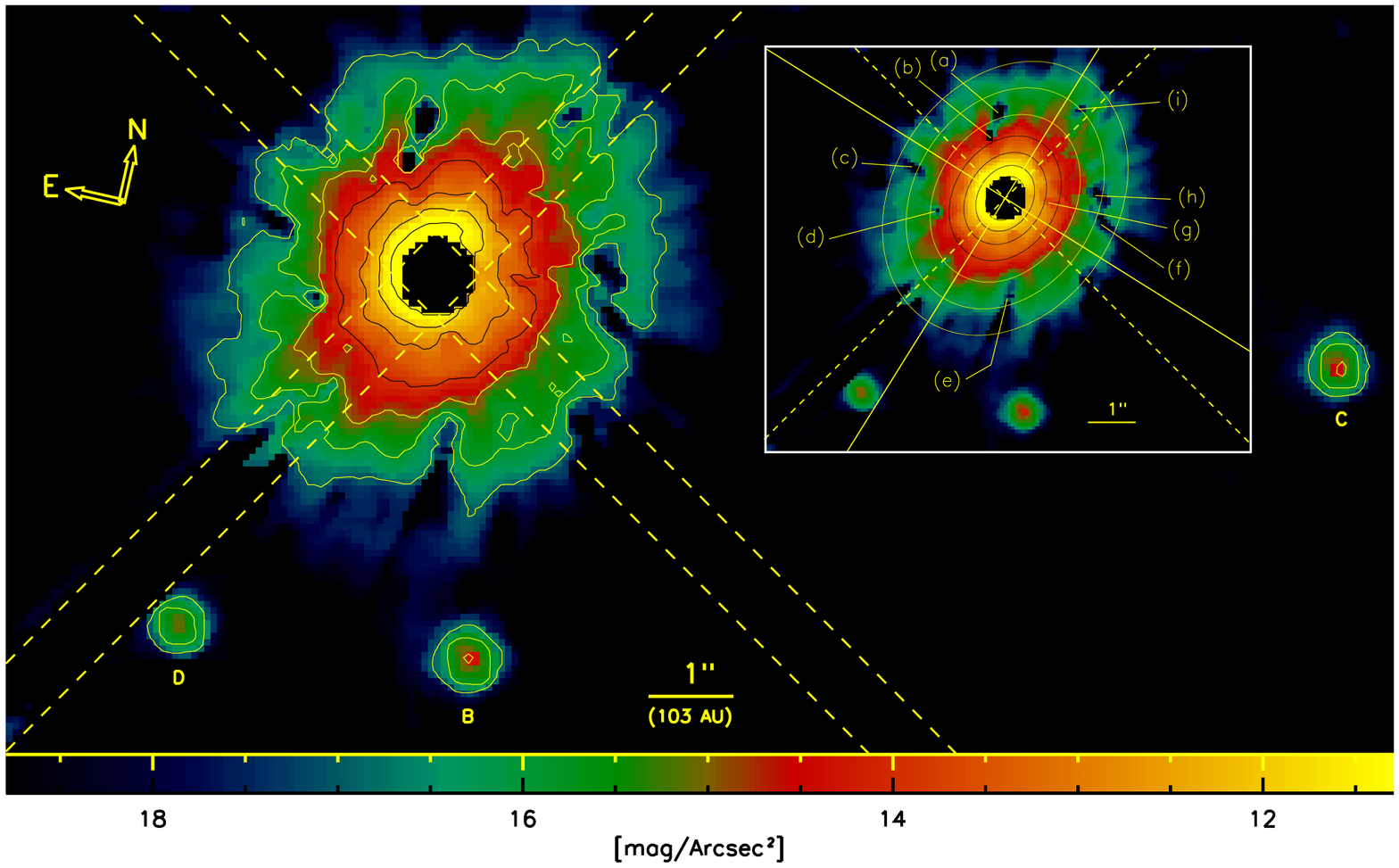}
\caption[]{Reduced image of the \hda\ disk observed in \co\ mode
  at 1.6\,$\mu$m with the HST/NICMOS2 camera. The \co\ mask is
  0.7$\arcsec$ in diameter ($\sim$\,9 pixels) whereas the numerical
  mask in the Figure is 0.9$\arcsec$. Isophotes from
  11\,mag/Arcsec$^2$ to 17\,mag/Arcsec$^2$ are indicated. The dashed
  lines give the width and the position of the spider arms. The small
  panel in the upper right corner shows the result of isophotes
  ellipse fitting.  Inhomogeneities due to the detector or to
  secondary spider diffraction spikes (which produce some kinds of
  fingers) are responsible for black spots. The most noticeable are
  labeled ($a$) to ($i$). In the same small panel, the plain lines
  represent the major and minor axis of the observed structure and the
  dashed ones represent the position of the spider arms.  }
\label{hd100disk}
\end{center}
\end{figure*}
\subsubsection{Resolved disk and vicinity}
The azimuthally averaged radial profile of the ratio between \hda\ and
the reference star \psfhda\ shows two basic distinct regimes: up to
3.5--3.8$\arcsec$, the profile decreases with distance from the star
then reaches a plateau further out. The same behavior arises using the
A2 and A1 reference stars dedicated to \mwc\ and HD\,141569
respectively. A resolved structure around \hda\ is then detected below
3.5--3.8$\arcsec$. A 3$\sigma$ uncertainty of 2\% is obtained on the
scaling factor used to subtract the reference star to \hda\ which
directly impacts on the precision of photometric measurements.

The final calibrated image is shown in Figure \ref{hd100disk} and
reveals an extended and elliptical structure centered on the star.  We
interpret this structure in terms of \cs\ material located in an
inclined disk. A visual inspection suggests the direction subtended by
the spider arm at a position angle (PA, measured east of north) of
$\sim$ 148$\degr$ as an axis of symmetry for the inclined disk, at
least at large distances. A more robust method (ellipse-fitting to the
isophotes) leads to a PA of 161$\degr\pm$5$\degr$ but the result might
be affected by different noise sources. Whatever the method, the
global direction of the disk is consistent with \citet{pan00}
observations but the precise PA is found to be at least 20$\degr$
larger than the measurements of \citet{pan00}. The axis ratio
corresponds to a disk inclination of 51$\degr\pm$3$\degr$ with respect
to the line of sight, assuming that the disk is axisymmetric, and
agrees with the results of \citet{pan00}.  We measure a total flux
density of 73$\pm$7\,mJy (the uncertainty is dominated by the
uncertainty on the scaling factor used to subtract the PSF star).

Some companion stars are also detected in the close vicinity of \hda.
The three brightest ones are shown in Figure \ref{hd100disk} and
measured astronomical parameters, namely projected separations, PAs
and magnitudes, are summarized in Table \ref{hd100companions}.
\begin{table}[!h] 
\begin{center}
\caption{\label{hd100companions}
  Projected distances, position angles (PAs) and magnitudes in filter
  F160W ($\lambda\dma{c}\simeq1.6\,\mu$m) of the 3 brightest visual
  companion stars detected in the vicinity of \hda. The labels refer
  to Figure \ref{hd100disk}.  3$\sigma$ uncertainties on the positions
  and PAs are mainly due to the uncertainty on the central star
  position below the mask but also to the uncertainty on the companion
  star centroid. The 3$\sigma$ photometric uncertainty is induced by
  the uncertainty on the scaling factor used to subtract the PSF star
  to the star of interest (subsection \ref{reduction}).}
\begin{tabular}[h]{cccc}
\hline
Label & Projected distance& PA & magnitude \\
 & {\scriptsize [Arcsec]}& {\scriptsize [degrees]} & {\scriptsize filter F160W} \\
\hline
B & $4.50\pm 0.15$ & $197.7\pm 1.1$ & $15.74\pm 0.04$ \\
C & $10.80\pm 0.15$ & $277.1\pm 0.5$ & $15.79\pm 0.11$ \\
D & $5.09\pm 0.15$ & $156.3\pm 0.9$ & $16.09\pm 0.08$ \\
\hline
\end{tabular}
\end{center}
\end{table}
\begin{figure*}[tbph]
\hbox to \textwidth
{
\parbox{0.50\textwidth}{
\includegraphics[angle=90,width=0.49\textwidth,origin=br]{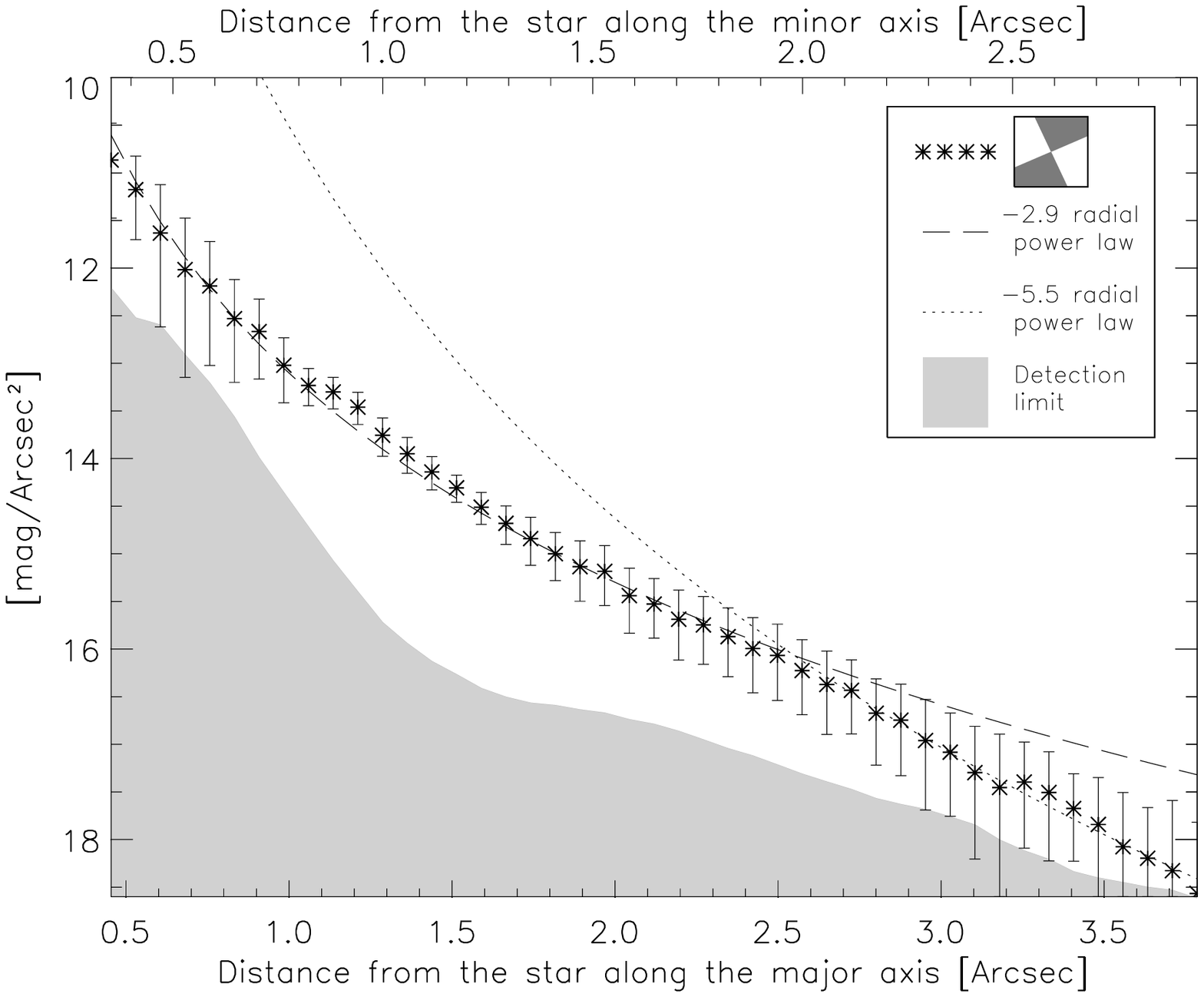}
}
\hfil
\parbox{0.50\textwidth}{
\includegraphics[angle=90,width=0.49\textwidth,origin=br]{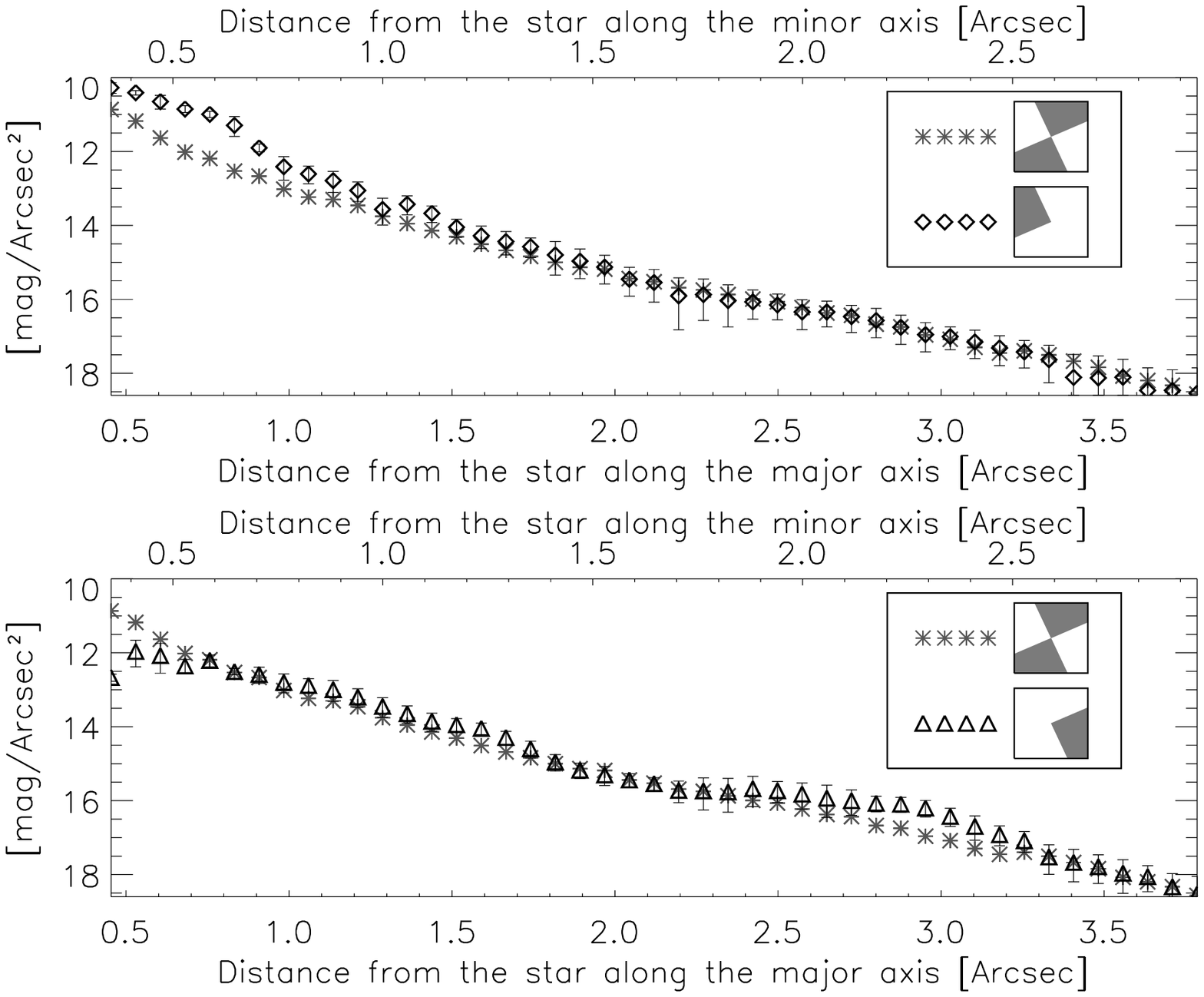}
}
}
\caption[]{ The \hda\ disk radial surface brightness profiles azimuthally
  averaged on elliptical contours assuming PA = 161$\degr$ for the
  major axis and an eccentricity of $e=0.63$. The left panel shows the
  major axis profile and the right ones the minor axis profiles.  The
  dark grey areas in the small boxes in the legend mimic the angular
  sectors centered on the star used in each case. These results are
  discussed in paragraphs \ref{discuss_profil} and
  \ref{discuss_asymmetry}.}
\label{profil}
\end{figure*}
\subsubsection{Radial surface brightness profiles}
\label{discuss_profil}
The radial brightness profiles of the disk are assessed by azimuthally
averaging the surface brightness on elliptical contours. The radial
profile measured towards the major axis of the disk is shown in the
left panel of Figure \ref{profil}. It reveals a smooth and continuous
decrease of the surface brightness with the distance from the star
from the very close edge of the \co\ mask up to 2.5$\arcsec$.  Beyond
2.5$\arcsec$, the profile is steeper and reaches the detection limit
(grey area in Figure \ref{profil}) at 3.5--3.8$\arcsec$. Basic radial
power-laws properly fit this measured surface brightness profile with
indexes $-2.92\pm 0.04$ in the radial range
[0.5$\arcsec$,2.5$\arcsec$] and $-5.5\pm 0.2$ outside of 2.7$\arcsec$.

The surface brightness profiles along the minor axis (NE and SW sides
of the disk) are shown in the right panels of Figure \ref{profil}. To
allow direct comparison with the results along the major axis, we plot
the profiles versus deprojected distances. They correspond to the
observed distances along the minor axis over $\sin(51\degr)$. As a
confirmation of the isophote ellipse fitting procedure, the main shape
of the minor axis profiles superimposes well on the major axis one.
\subsubsection{NE-SW brightness asymmetry}
\label{discuss_asymmetry}
However, a significant brightness enhancement arises in the NE
direction mainly below 0.7--0.8$\arcsec$ in projected distances but
not in the opposite direction. Also in the same radial range, the
behavior of the SW profile better follows the major axis profile down
to $\sim$0.5$\arcsec$. Below this distance, the SW side of the disk is
slightly fainter but with a poor level of confidence (very close to
the detection limit). This NE-SW brightness asymmetry is clearly
evidenced in Figure \ref{hd100disk}.

\begin{figure}[tbp]
\begin{center}
\hspace*{1.cm}
\includegraphics[angle=90,width=0.4\textwidth,origin=br]{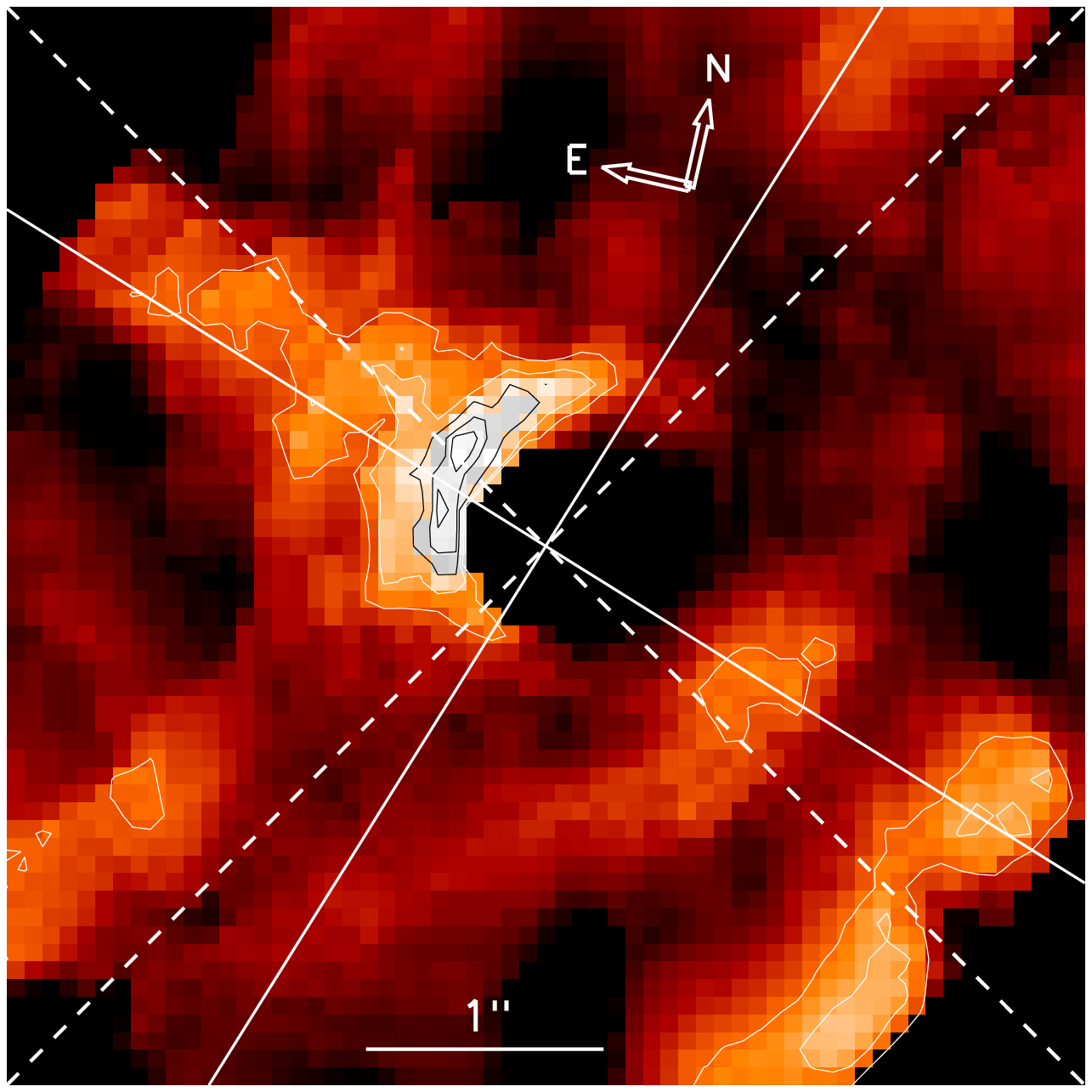}
\vspace{1truemm}\par\noindent
\includegraphics[angle=90,origin=bl,width=0.49\textwidth]{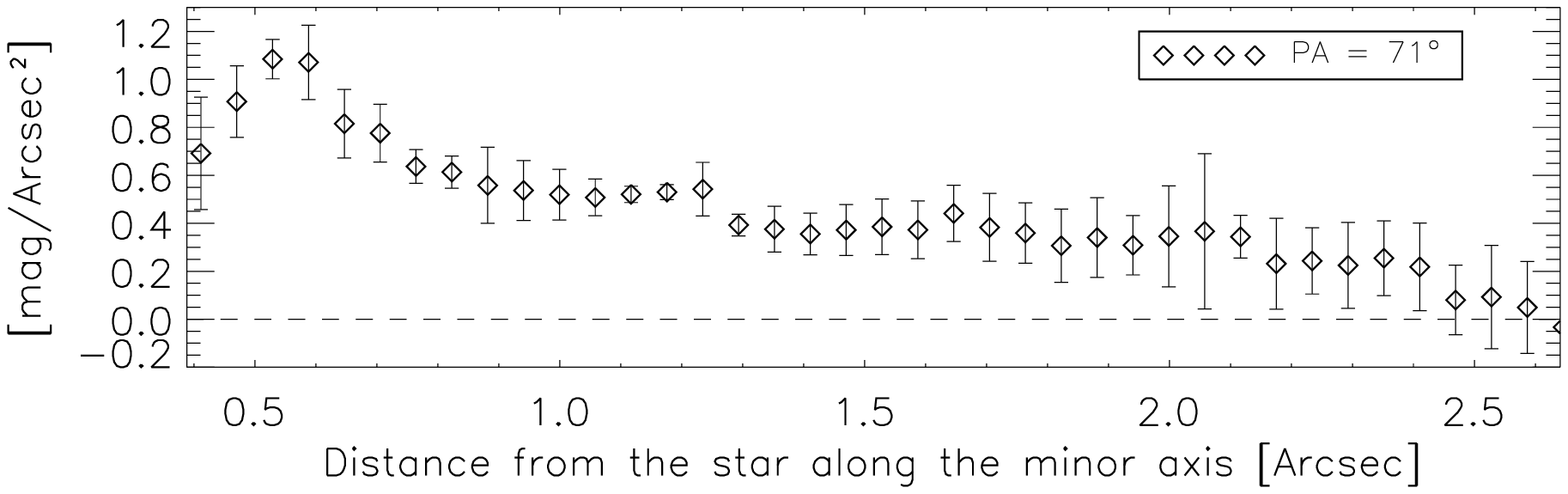}
\caption[]{ \textit{Upper panel}~: Residue resulting from the subtraction
  of a synthetic disk with an axisymmetrical surface brightness
  strictly identical to the one shown in the left panel of Figure
  \ref{profil} and inclined at 51$\degr$ from edge-on towards
  PA=161$\degr$ to the reduced image shown in Figure \ref{hd100disk}.
  For each pattern, the lowest level is 0.4 mag/Arcsec$^2$.  The white
  contours are spaced at 0.4 and 0.6 mag/Arcsec$^2$, the black ones at
  0.8, 1 and 1.1 mag/Arcsec$^2$. As for Figure \ref{hd100disk}, the
  plain lines represent the major and minor axis and the dashed ones
  the position of spider arms.  \textit{Lower panel}~: radial profile
  along the NE minor axis averaged for each distance from the star
  over 9 pixels in the perpendicular direction.  The subtraction
  highlights the NE-SW brightness asymmetry. It allows to quantify
  this asymmetry and to determine the PA of the excess.}
\label{exces}
\end{center}
\end{figure}

To emphasize the asymmetry, we subtract a synthetic disk which fits
the disk isophotes to the observed disk of \hda. In the NE side, the
subtraction shown in Figure \ref{exces} reveals an excess of flux up
to 1.1\,mag/Arcsec$^2$ at 0.5$\arcsec$-0.6$\arcsec$ collimated in a
direction close to the minor axis of the disk (plain line). In the SW
side, the systematic pattern at about 2.2$\arcsec$, responsible for
the shape of the radial profile shown in the bottom right panel of
Figure \ref{profil}, is clearly evidenced.

We first checked whether this asymmetry could result from a bad
centering of the PSF star during the reduction procedure. To
significantly decrease the inner brightness asymmetry, the reference
star has to be shifted by more than 1.5 pixel in the NE direction.
This leads to unrealistic strong asymmetries in all the rest of the
image which characterize an obvious bad centering of the subtracted
PSF star. The asymmetry arises close to the position of a diffraction
spike which might seem suspicious. But due to a good alignment, the
diffraction spikes were completely removed during PSF subtraction and
did not require any further process which would have directly cause
the effect.

Could it be due to instrumental effects then? \hda\ and the reference
star have been observed during the same orbit. Therefore,
orbit-to-orbit variations, such as the migration ($\sim$ 0.25 pixel)
of the coronagraphic hole on the detector or defocus for instance
\citep{sch99}, cannot account for the observed effect.  Another issue
concerns the mis-centering of the star behind the mask during target
acquisition.  An accuracy of a few tenths of pixel is reached when
centering PSF on the occulting mask but a mis-centering of the star by
only a third of pixel can produce brightness asymmetries mainly within
0.4$\arcsec$--0.5$\arcsec$ in radius \citep{sch99}.  Observations at
different roll angles would help to check whether the asymmetry is
real or an artifact.
\subsection{Discussion and interpretation}
\subsubsection{Disk radiation to stellar luminosity ratio at 1.6\,$\mu$m}
The measured flux density of the disk at 1.6\,$\mu$m corresponds to
10.4$\pm$0.1\,mag according to the zero point flux provided by the
STScI.  On the other hand, the V-H color index observed by
\citet{hu89} and more recently Kurucz spectrum fitting by
\citet{mal98} indicate that the disk emission represents 55\% to 60\%
of the total flux in the H-band.  This excess may be due to the
thermal emission of hot grains at very short distances from the
central star, presumably close to the grain sublimation limit (a few
fraction of AU, see also Figure \ref{temp}).  Obviously this excess
falls below the \co\ mask used during present observations (the mask
radius is 0.35$\arcsec$ or $\sim$36\,AU {\it in situ} according to
Hipparcos measurements). Assuming H=5.88 for \hda\ \citep{mal98}, our
flux density measurement provides a lower limit of $4\times 10^{-2}$
for the scattered to photospheric flux ratio at 1.6\,$\mu$m with an
uncertainty of $15$\%. This is about 2.5 less than the ratio inferred
by \citet{pan00} in J band.
\subsubsection{Normal optical thickness of the disk}
\label{thick}
\begin{figure}[tbph]
\includegraphics[angle=90,width=0.49\textwidth,origin=br]{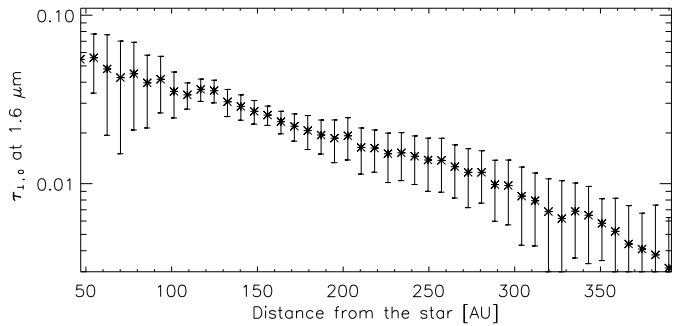}
\caption[]{1.6\,$\mu$m normal optical depth derived from the
  observed radial surface brightness distribution of the \hda\ disk in
  the optically thin approach. Between 50 and 100\,AU, it is a factor
  of between 40 and 50 less than the normal optical thickness inferred
  by \citet{pan00} at 1.25\,$\mu$m (J band).}
\label{hd100profopt}
\end{figure}

As a first step in analyzing the data, we assume that the disk is
optically thin whatever the direction. Since the disk is inclined with
respect to the line-of-sight, it is then straightforward to derive the
normal optical thickness $\tau_{\perp}(r)$ at 1.6\,$\mu$m. This
requires some reasonable assumptions on the optical dust properties.
Both method and assumptions are summarized in Appendix A. The result
is shown in the upper panel of Figure \ref{hd100profopt}. In this
simple approach, the disk appears indeed optically thin vertically at
1.6\,$\mu$m everywhere in the resolved radial range but $\tau_{\perp}$
reaches a quite high maximum of $\sim$ 0.06 close to the mask's edge.

Following \citet{kri00}, we then assess the midplane optical depth by
assuming that the vertical shape of the disk is a Gaussian with a
similar scale height of 20\,AU at 140\,AU from the star and a $\beta$
disk flaring index of $1$. We find a $0.15$ midplane optical depth
between 50\,AU and the outer edge of the resolved disk. It falls below
0.1 between $\sim$ 80\,AU and the outer edge. This would indicate that
the optically thin regime may not apply in the inner regions of the
disk and that part of the stellar light is obscured at large distances
in the midplane leading to an underestimation of $\tau_{\perp}$. This
is in line with the large millimeter flux \citep{hen94,hen98}, that
suggests that the disk is at least marginally optically thick in part.
\subsubsection{Normal surface density and minimum mass of the dust disk}
\begin{figure}[tbph]
\includegraphics[angle=90,width=0.49\textwidth,origin=br]{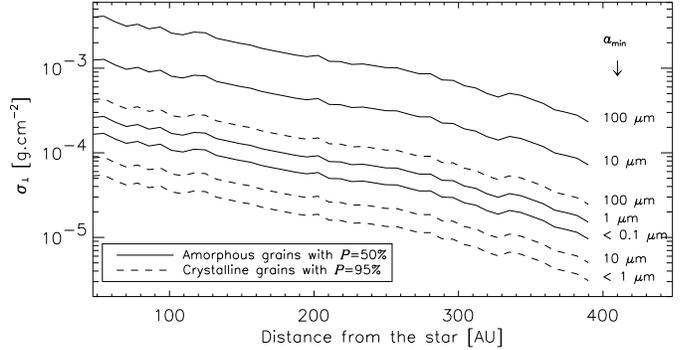}
\caption[]{Normal
  surface densities of the disk which reproduce, in the optically thin
  approach, the scattered light observations for different minimum
  grain sizes and the two types of grains considered. Note that
  surface densities as well as masses should be scaled by a factor
  $\sqrt{a\dma{max}\textrm{[cm]}}$ if the maximum grain size was not
  1\,cm as assumed here.}
\label{hd100surfdens}
\end{figure}

The optically thin approach is usefull to determine a lower limit on
the normal surface density $\sigma_{\perp}(r)$ in solids (see Appendix
B).  Since we assume a grain size distribution proportional to
$a^{-3.5}$, $\sigma_{\perp}(r)$ depends both on the minimum
($a\dma{min}$) and maximum ($a\dma{max}$) grain sizes in the disk.
Dust grains are assumed to be icy porous aggregates made of a silicate
core and coated by organic refractories. Two dust chemical
compositions are considered: amorphous and crystalline grains with
porosities of 0.5 and 0.95 respectively \citep[see appendix A and][for
more details on the model used]{aug99a}. $\sigma_{\perp}(r)$ is
plotted in Figure \ref{hd100surfdens} for different minimum grain
sizes.  Between 50\,AU and 100\,AU for instance, $\sigma_{\perp}(r)$
is as large as a few $10^{-5}$\,g.cm$^{-2}$ and about one order of
magnitude larger if grains are at the blow-out size limit ($\sim
13\,\mu$m, see Figure \ref{albedo}). In terms of dust mass between
50\,AU and 390\,AU, this corresponds to a few Earth masses at most
(the dust mass is about $3750 \times \sigma_{\perp}(50\mathrm{AU})\,
$M$_{\oplus}$) and is at least one order of magnitude less than dust
masses inferred from infrared or millimeter excesses
\citep{bou00,hen98}. Obviously, if meter-sized bodies are formed, the
normal surface density as well as the mass in solids should be
multiplied by at least a factor of 10. The dust mass may also be
significantly larger if the optical thickness inferred is
underestimated.
\subsubsection{Shape of the dust distribution in the optically
  thin approach}
$\sigma_{\perp}(r)$ follows a $-0.92\pm 0.04$ radial power law up to
about 270\,AU.  Similar distributions are found for disks around
T\,Tauri stars \citep[e.g.][]{dut96}. The modeling of cloud core
infall with no viscosity also predicts disk surface densities
proportional to $r^{-1}$ \citep{lin90}. Similarly, the surface density
of the solar nebula during the disk dissipation is expected to follow
a $r^{-1}$ radial power law under the assumption that the utilization
of available solids to built planets in the Solar System is not
efficient \citep{cam95}. In that frame, and given the conclusions on
the optical thickness in section \ref{thick}, the inner part of the
disk can rather be considered of first generation.

Outside of 270\,AU, $\sigma_{\perp}(r)$ falls as $r^{-3.5\pm 0.2}$. A
similar discontinuity of the dust distribution has already been
reported in the case of the \bp\ disk. If the dynamics of the \hda\ 
disk further than 270\,AU is similar to that of the outer part of \bp\ 
disk (namely, supplied in small grains by collisions among
planetesimals and radiation pressure), we would expect the disk radial
density distribution to follow a $-4$ radial power law \citep{lec96}.
Such a disk, assumed optically thin and seen perfectly edge-on, indeed
produces a midplane surface brightness proportional to $r^{-5}$
\citep{nak90} as typically observed for \bp\ \citep{hea00}. The latter
conclusion also requires quite isotropic scattering properties. With
these assumptions, the \hda\ disk would then flare as $r^{0.5\pm
  0.2}$. But the dynamical similarities between the \hda\ and \bp\ 
disks at large distances remains an open issue.
\subsubsection{On the NE-SW brightness asymmetry}
What can produce the NE-SW brightness asymmetry if this is not an
artifact due to instrumental effects? Although dust distribution
asymmetries can not be ruled out, the fact that the brightness
asymmetry of the \hda\ disk occurs in a direction close to its minor
axis tends to indicate that this would be rather due to scattering
properties.

In the case of an optically and a geometrically thin inclined disk,
anisotropic scattering properties can produce similar asymmetries but
they should occur whatever the distance from the star. In present
case, the asymmetry mostly appears at short distances where the normal
optical thickness reaches values close to $0.1$.

The transition between the (marginally) optically thick regime in the
inner disk and the optically thin regime at large distances combined
with anisotropic scattering properties may explain the shape of the
observed NE-SW brightness asymmetry. Further observations and models
are needed to confirm that issue.
\subsubsection{Dust temperature and minimum grain size in the disk}
\begin{figure}[tbph]
\includegraphics[angle=90,width=0.49\textwidth,origin=br]{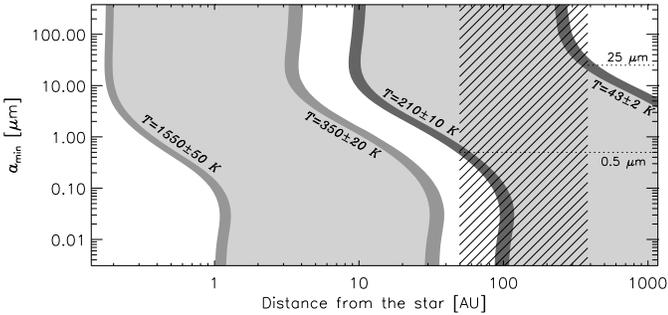}
\caption[]{Four mean isotherms of amorphous grains versus the distance
  from the star and the minimum grain size in the disk. The dark
  (resp.  grey) isotherms correspond to the inner and outer boundaries
  of the cold (resp. hot) dust population proposed by \citet{mal98}
  from ISO spectra. The dashed area indicates the radial range in
  which the disk is detected in scattered light.}
\label{temp}
\end{figure}

\citet{hu89} first claimed that the \hda\ dust disk consists of a cold
and a hot dust populations to account for the overall IR excesses. The
relatively high spectral resolution of ISO spectra between
$\lambda=2.4\,\mu$m and 180\,$\mu$m yielded \citet{mal98} to propose
temperature ranges for the two dust populations. Figure \ref{temp}
shows these four isotherms as a function of the distance from the star
and of the minimum grain size $a\dma{min}$ in the disk. The light-grey
areas represent then the two dust populations. We only consider
amorphous grains since amorphous silicates are supposed to be at least
ten times more numerous than crystalline ones \citep{mal98}.

In terms of spatial extent, the cooler population derived by
\citet{mal98} is consistent with the dust resolved in scattered light
if the grains are larger than $\sim 0.5\,\mu$m (see Figure \ref{temp}:
intersection of the light-grey and dashed areas). Indeed, grains at
$T=210\pm 10$\,K (the temperature corresponding to the inner edge of
the colder population) smaller than a half micron lie further than
$\sim 50$\,AU. Such grains would induce a detectable cut-off in the
dust distribution somewhere between 50\,AU and 100\,AU which is not
observed.
\subsubsection{Consistency of the scattered light image with IR excesses}
The dust resolved in scattered light, colder than $\sim$\,210\,K
according to \citet{mal98}, is expected to be responsible for most of
the long wavelength IR excesses. At 100\,$\mu$m, the SED is mainly
featureless and the IR excess mostly comes from an underlying
continuum \citep{mal98}.  We can then compare the observed flux at
100\,$\mu$m to the thermal emission implied by the normal surface
densities deduced from present scattered light images in the optically
thin approach (Figure \ref{hd100profopt}).
\begin{figure}[tbph]
\includegraphics[angle=90,width=0.49\textwidth,origin=br]{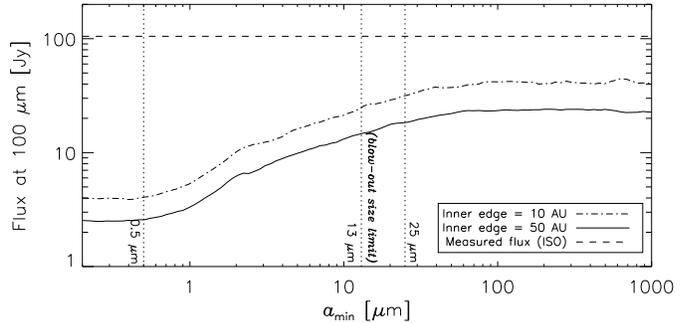}
\caption[]{Contribution to the $\lambda=100\,\mu$m thermal emission of
  the dust population resolved in scattered light down to 50\,AU as a
  function of the minimum grain size $a\dma{min}$ (plain line).  The
  dotted-dashed line corresponds to the same population but
  extrapolated down to 10\,AU. The uncertainties on this estimation
  are: 1/ the $\sim$\,10\% uncertainty on the scattered image
  photometry, 2/ the uncertainty on the anisotropic scattering
  properties (we have assumed $f(90\degr)=1/4\pi$ for simplicity, see
  Appendix A) 3/ all the molecules detected with ISO which could
  partially change the mean grain albedo and 4/ the exact surface
  density if the optical depth is underestimated.}
\label{flux100um}
\end{figure}

Figure \ref{flux100um} shows this comparison versus $a\dma{min}$ for
two different disk inner edges: the first one at 50\,AU and the second
at 10\,AU. In the latter case, $\sigma_{\perp}(r)$ has been
extrapolated down to 10\,AU with a -0.92 radial power law. The
distance of 10\,AU has been chosen because it corresponds to the
position of 210\,K grains larger than typically the blow-out size
limit ($\sim 13\,\mu$m, Figures \ref{albedo} and \ref{temp}).

Whatever $a\dma{min}$, the predicted flux never reaches the observed
limit. The optically thin approach does not then bring any upper limit
on the minimum grain size in the disk. Even grains larger than
10--20\,$\mu$m lying in the 10\,AU-inner-edge disk radiate a flux 2.5
to 3 times smaller than the measured emission.

As already noticed, the optically thin approach may underestimate the
surface density, indicating that part of the dust mass does not
contribute to scatter light images but may emit at 100$\,\mu$m. For
instance, with the assumed dust distribution, increasing the dust mass
by a factor of 40 would still be compatible with the observations if
the grains are submicronic.
%
\section{\hdb}
\label{hd135}
\subsection{The HD\,135344 multiple system}
The F4Ve star \hdb\ \citep{wae90a,wae90b} forms a visual binary system
named HD\,135344 with \psfhdb\ $\sim$20.4$\arcsec$ away almost in the
Northern direction. The latest star, that we used as the comparison
star during our HST/NICMOS2 observations, has a spectral type between
F2 (CDS) and A2 \citep{mye98}.  Among our targets, \hdb\ is the less
luminous star but exhibits the largest relative infrared excess
\citep[$L\dma{IR}/L_*=0.67$,][]{cou95}. Dust emission features at
3.29\,$\mu$m, 7.8\,$\mu$m and 11.3\,$\mu$m have been evidenced by
\citet{cou95} and attributed to aromatic molecules of a few
angstr\"oms \citep{syl96}. CO is also detected by \citet{zuc95} and
\citet{cou98} whose abundance suggests a low gas to dust ratio
\citep[$8\times 10^{-3}$,][]{cou98} if CO is representative of the gas
content (i.e. not depleted with respect to the other species)
consistent with that of Vega-like stars.

\citet{cou95} and \citet{mal98} derived from the full SED modeling a
two dust component description of the disk. However, without imaging,
the cooler dust that is present in the outer part of the disk ($>$
typically 10 AU) is not precisely constrained.  Moreover, as shown by
\citet{ken97}, a single thin (not flared) disk can not account for the
observed infrared excess and yields \citet{syl97} to propose either
the presence of an envelope or accretion activity.
\subsection{Results}
\begin{figure}[tbp]
\begin{center}
\includegraphics[angle=90,origin=bl,width=0.49\textwidth]{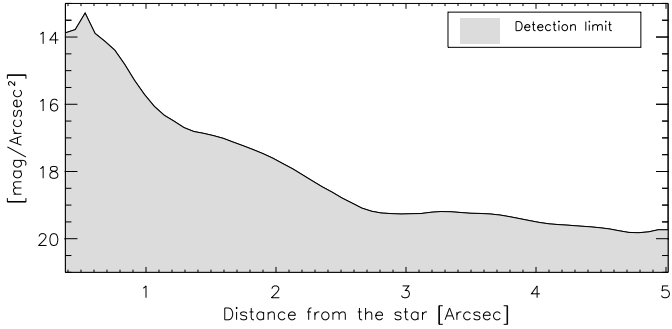}
\caption[]{\hdb\ detection limit at 1.6$\,\mu$m with the HST/NICMOS2 camera.}
\label{hd135detlim}
\end{center}
\end{figure}
\subsubsection{Disk detection limit}
\begin{figure}[tbp]
\begin{center}
\includegraphics[angle=90,origin=bl,width=0.49\textwidth]{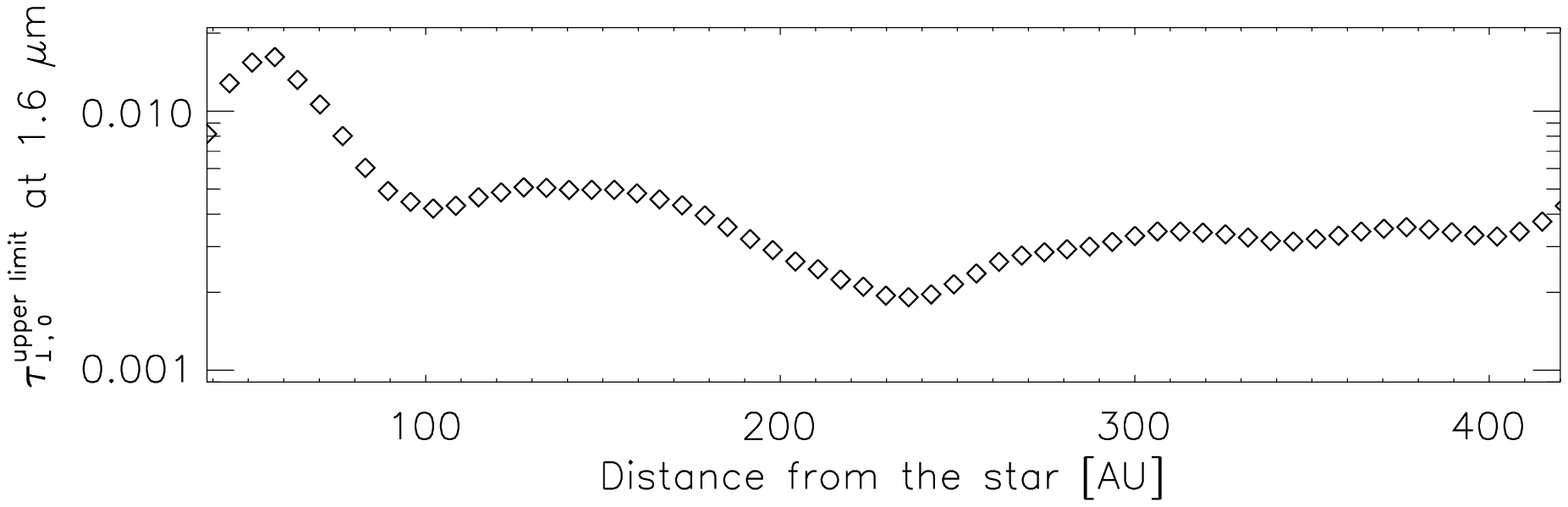}
\vspace{1truemm}\par\noindent
\includegraphics[angle=90,origin=bl,width=0.49\textwidth]{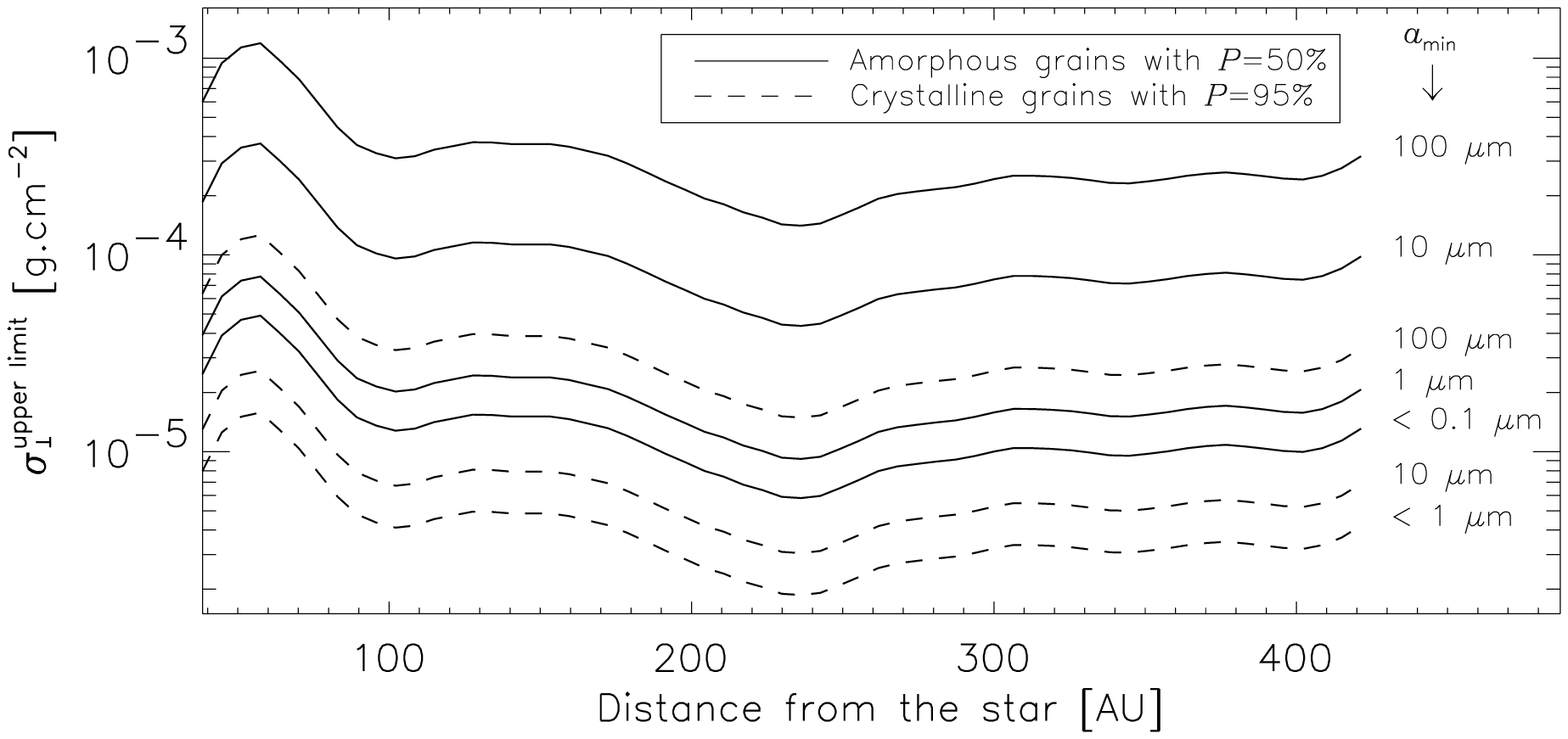}
\caption[]{Upper limits on the disk normal optical thickness at
  $\lambda=1.6\,\mu$m (upper panel) and on the dust surface density as
  a function of the minimum grain size $a\dma{min}$ (lower panel) for
  \hdb\ assuming that the disk is optically thin.}
\label{hd135profopt}
\end{center}
\end{figure}

No obvious feature that could be attributed to \cs\ dust is detected
in the 1.6\,$\mu$m images of \hdb\ (Figure \ref{hd135double}). The
corresponding detection limit (Figure \ref{hd135detlim}) shows that at
1.6\,$\mu$m the disk is fainter than 13.5--14\,mag.arcsec$^{-2}$
further than 0.4$\arcsec$ ($\sim$35\,AU) and fainter than
16\,mag.arcsec$^{-2}$ at 93\,AU (1.1$\arcsec$), {\it i.e.} close to
the inner disk edge derived by \citet{syl97} from SED modeling.  In
the optically thin approach, this can be interpreted in terms of an
upper limit on the normal optical thickness of about $10^{-2}$ (Figure
\ref{hd135profopt}) outside of 0.4$\arcsec$ in radius (35\,AU).  The
dust material further than 35\,AU (if any) then is either optically
thin at 1.6\,$\mu$m or does not contribute at this wavelength due
starlight occultation by the dust closer to the star.
\subsubsection{Implications on the dust distribution}
With the rough assumption that the dust surface density is
proportional to $r^{-1.2}$ (as typically observed for young stars and
which implies the larger mass), the dust mass between 35\,AU and
420\,AU would be less than: $2270 \times
\sigma_{\perp\,,\,r=35\mathrm{AU}}\uma{upper\,limit}\, $M$_{\oplus}$.
It corresponds to $\sim$0.25\,M$_{\oplus}$ assuming
$\sigma_{\perp\,,\,r=35\mathrm{AU}}\uma{upper\,limit}\sim
10^{-4}$\,g.cm$^{-3}$. This must be compared to the few Earth masses
required to reproduce the full SED \citep{syl97,cou98} assuming
moreover a maximum grain size at least 10 times smaller than for
present work.

A large fraction of the dust mass (optically thin approach) or at
least a large amount of dust (optically thick approach) is then
certainly confined within the first few tens of AU. This is consistent
with the mean cold population position of 23\,AU derived by
\citet{cou95} but not with \citet{syl97} (who assume very small grains
in the disk, $\sim$ 50\,\AA).  Present results agree with the disk
size limit of $\sim$\,60\,AU inferred by \citet{jay00} from their
unresolved 10 and 18\,$\mu$m images.
\subsubsection{A tight binary system close to \hdb}
\begin{figure}[tbph]
\begin{center}
\includegraphics[angle=90,width=0.49\textwidth,origin=br]{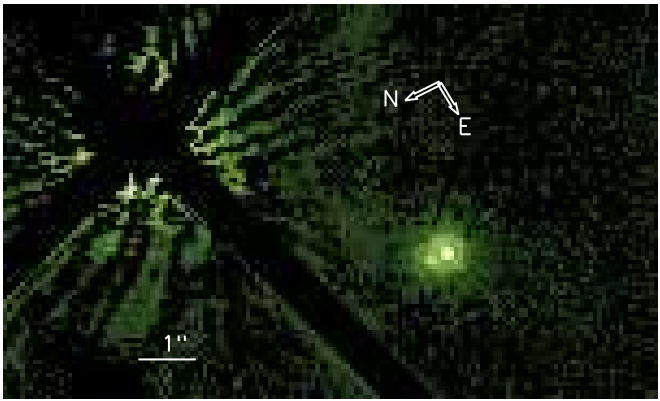}
\caption[]{The close binary system detected in the vicinity of \hdb\ at
  1.6\,$\mu$m with the HST/NICMOS2 camera.}
\label{hd135double}
\end{center}
\end{figure}

A close binary system in the SW direction is detected in the vicinity
of \hdb\ (Figure \ref{hd135double}). The brightest companion is
5.8$\arcsec\pm$0.15$\arcsec$ away from \hdb\ at
PA=129.8$\degr\pm$0.8$\degr$. The two close companion stars are
separated by only 0.32$\arcsec\pm$0.04$\arcsec$ almost aligned along
the NS axis; we find a PA of 352$\degr\pm$7 for the third companion
with respect to the brightest one. In Filter F160W, the secondary
companion has magnitude of 15.2$\pm$0.1 and the third companion is
about 6.3 times fainter (mag = 17.2$\pm$0.1).
%
%
\section{\mwc}
\label{mwc}
\subsection{A sub-arcsec dust disk}
Among our sample, \mwc\ is the only one for which \cs\ gas has been
imaged \citep{man97b}. The resolved structure shows Doppler shifts
characteristic of the disk rotation over angular scale of a few arcsec
and inclined by $\sim 60\degr$ from edge-on. Meanwhile, the dust
continuum emission in millimeter also appears elongated in the same
direction but is less extended (HWHM\,$\sim$\,0.6\,$\arcsec$ or
$\sim$\,80\,AU).

Two dust populations seem required to account for the overall infrared
excesses. For instance, a combination of a geometrically thin disk and
an envelope is proposed by \citet{mir99} \citep[see also][]{mal98}. As
in the case of $\beta$\,Pictoris, the 10\,$\mu$m spectrum of \mwc\ 
shows features attributed to silicates (in particular crystalline
olivine) similar to that of long-period comets in our solar system
\citep[e.g. comet Levy,][]{sit99}.

\subsection{Detection limit and implications on the dust distribution}
\begin{figure}[tbp]
\begin{center}
\includegraphics[angle=90,origin=bl,width=0.49\textwidth]{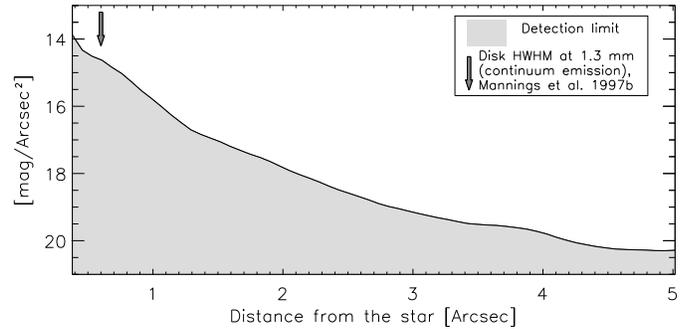}
\caption[]{\mwc\ detection limit at 1.6$\,\mu$m with the HST/NICMOS2 camera.}
\label{mwc480detlim}
\end{center}
\end{figure}

According to the detection by \citet{man97b}, the dust disk is
expected inside an angular region smaller than 1$\arcsec$ in radius, a
region where it is challenging to resolve faint structures because of
PSF subtraction residues and diffraction spikes for example.
Actually, very close to the edge of the mask, we do not find any
convincing features which would clearly reveal the presence of the
\cs\ disk. The azimuthally averaged noise on the image, interpreted in
terms of detection limit, is shown in Figure \ref{mwc480detlim}.

\begin{figure}[tbp]
\begin{center}
\includegraphics[angle=90,origin=bl,width=0.49\textwidth]{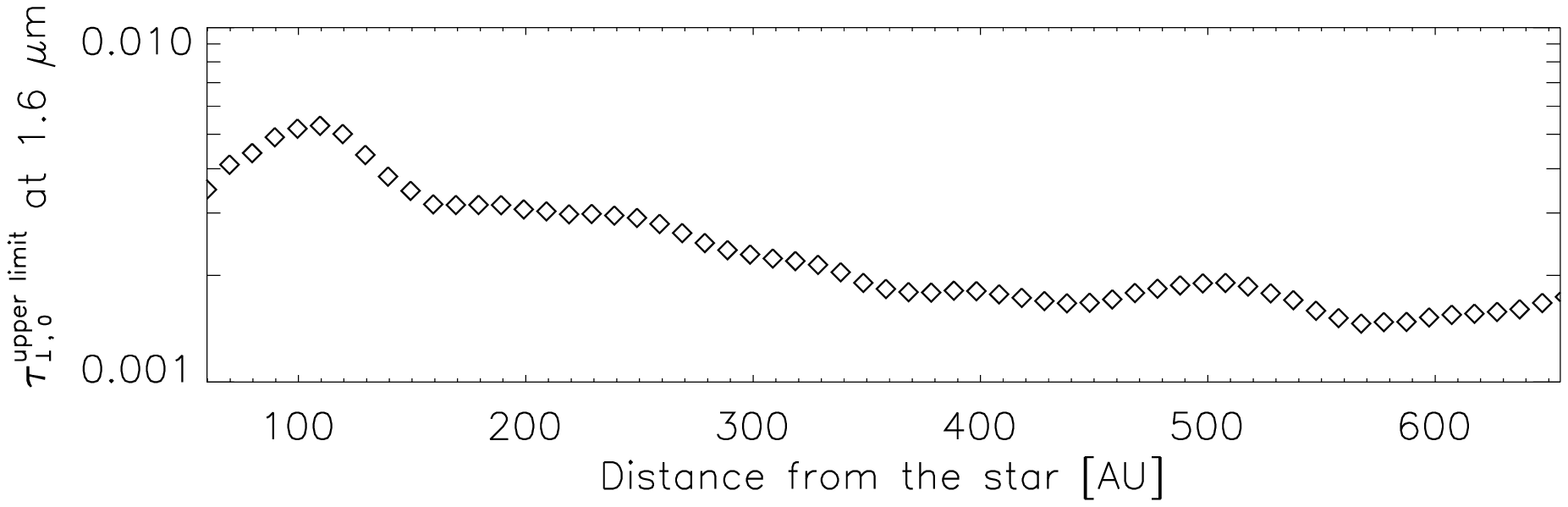}
\vspace{1truemm}\par\noindent
\includegraphics[angle=90,origin=bl,width=0.49\textwidth]{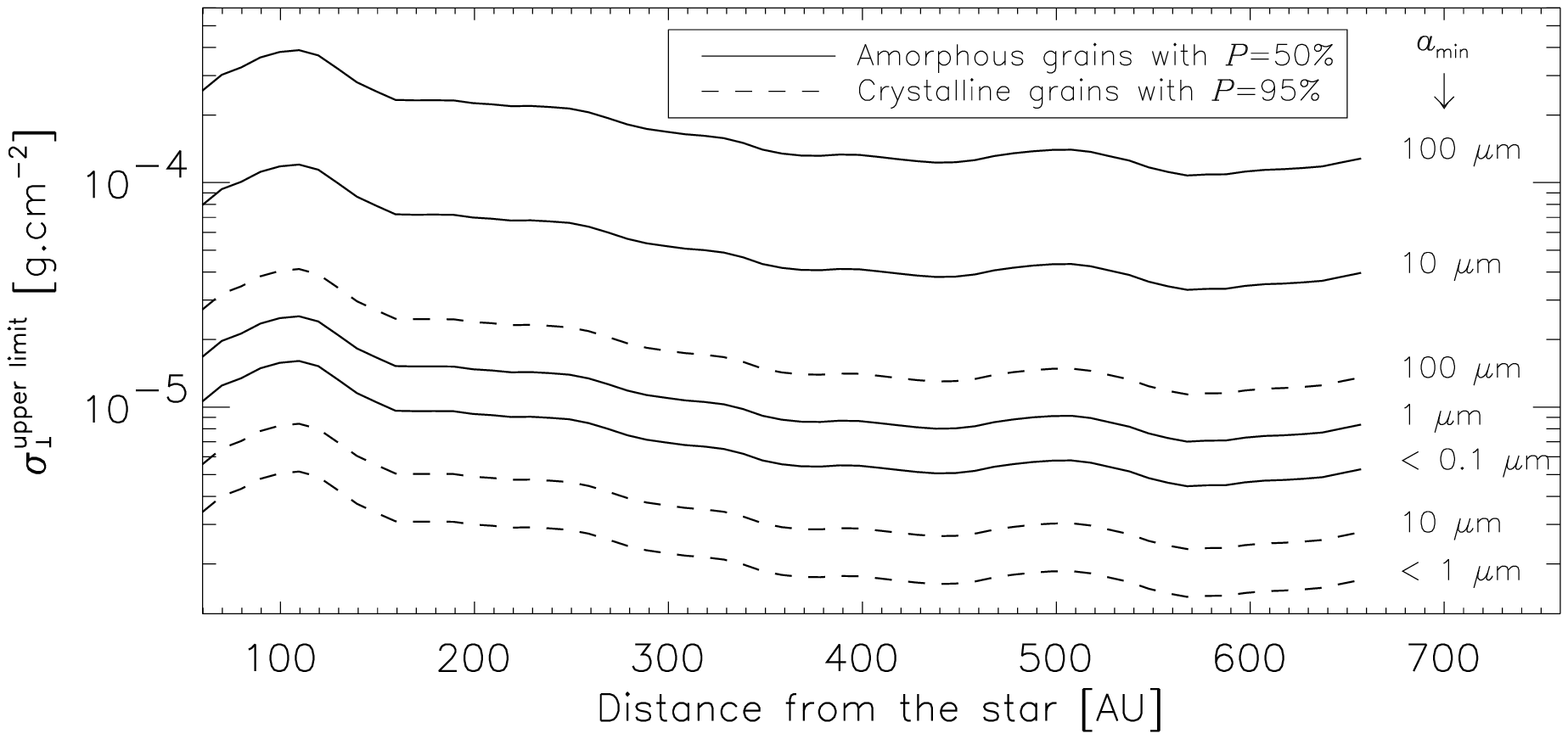}
\caption[]{Upper limits on the \mwc\ disk optical thickness at
  $\lambda=1.6\,\mu$m and on the dust surface density as a function of
  the minimum grain size $a\dma{min}$.}
\label{mwc480profopt}
\end{center}
\end{figure}

We applied in the case of \mwc\ strictly the same technique as the one
used to derive some interesting properties on the \hdb\ disk.
Similarly, we conclude that either the disk further than 60\,AU is
optically thin at 1.6\,$\mu$m or that the starlight is at least
partially masked by a large amount of dust inside 60\,AU (Figure
\ref{mwc480profopt}).

For instance, if we constrain the dust mass between 60\,AU and 660\,AU
to be smaller than 170\,M$_{\oplus}$ in solids and assuming
$\sigma_{\perp} \propto r^{-1.75}$ \citep{man97b}, we find an upper
limit on the surface density at 60\,AU of $6\times
10^{-2}$\,g.cm$^{-2}$. This is consistent with the results in the
optically thin approach from our scattered light observations (Figure
\ref{mwc480profopt}) even if grains are as large as a few hundred
microns and even if we rather assume that only 10\% of the dust mass
lies further 60\,AU. The present observations confirm then that most
of the the dust mass is confined inside the first hundred of AU as
shown by \citet{man97b}.
%
%
\section{Conclusion}
Among our four HST/NICMOS2 Cycle 7 targets, two \cs\ disks have been
resolved in scattered light. In this paper, we report the positive
detection of an extended inclined disk around \hda\, a typical
transient old-PMS star expected to be in a disk dissipating phase. The
shape of the derived surface density, the normal optical depth and, if
confirmed, the brightness asymmetry along the minor axis of the
inclined disk indeed indicate that the \hda\ \cs\ environment
certainly traces a stage between ``first generation'' massive disks
around PMS stars and those of ``second generation'' around MS stars
(Vega-like disks).

Disk non-detections are reported for both \hdb\ and \mwc. In each
case, the detection limit in scattered light allows to infer
constraints on the dust location consistent with observations at other
wavelengths.
\begin{acknowledgements}
  We thank the referee, Dr. M.L. Sitko, for helpful suggestions.
\end{acknowledgements}
\appendix
\section{Disk normal optical thickness}
The disk normal optical thickness can be straightforwardly derived
from the normal surface brightness if we assume that the disk is
axisymmetrical and optically thin.

More precisely, the disk normal surface brightness in scattered light is:
\begin{eqnarray*}
SB_{\perp}(r) \simeq \Phi_{\lambda} \, r^{-2} \, f(90\degr) \times \langle
\sigma\dma{sca}\rangle_a \, n_{\perp}(r)
\end{eqnarray*}
where: $\Phi_{\lambda}$ is the total received flux (star+disk) at the
considered wavelength $\lambda$, $r$ is the distance from the star,
$f$ is the scattering phase function, $\langle
\sigma\dma{sca}\rangle_a$ is the scattering cross-section averaged
over the grain size distribution and $n_{\perp}(r)$ is the grain
normal surface density distribution. This equation assumes that grain
properties do not strongly depend on $r$.

On the other hand, the normal optical thickness is given by:
\begin{eqnarray*}
\tau_{\perp}(r) = \langle \sigma\dma{ext}\rangle_a\,n_{\perp}(r)
= \omega^{-1}\times\langle \sigma\dma{sca}\rangle_a \, n_{\perp}(r) \\
\textrm{with:}\,\,\,\,\, \omega = \frac{\langle \sigma\dma{sca}\rangle_a}
{\langle \sigma\dma{ext}\rangle_a \,}
\end{eqnarray*}
where $\langle \sigma\dma{ext}\rangle_a$ is the averaged extinction
cross-section. Finally, we have:
\begin{eqnarray*}
\tau_{\perp}(r) \simeq \omega^{-1}\, \Phi_{\lambda}^{-1}  \, f(90\degr)^{-1} \,
r^2\,SB_{\perp}(r)\,\,\,\,\, .
\end{eqnarray*}

\begin{figure}[tbp]
\begin{center}
\includegraphics[angle=90,origin=bl,width=0.49\textwidth]{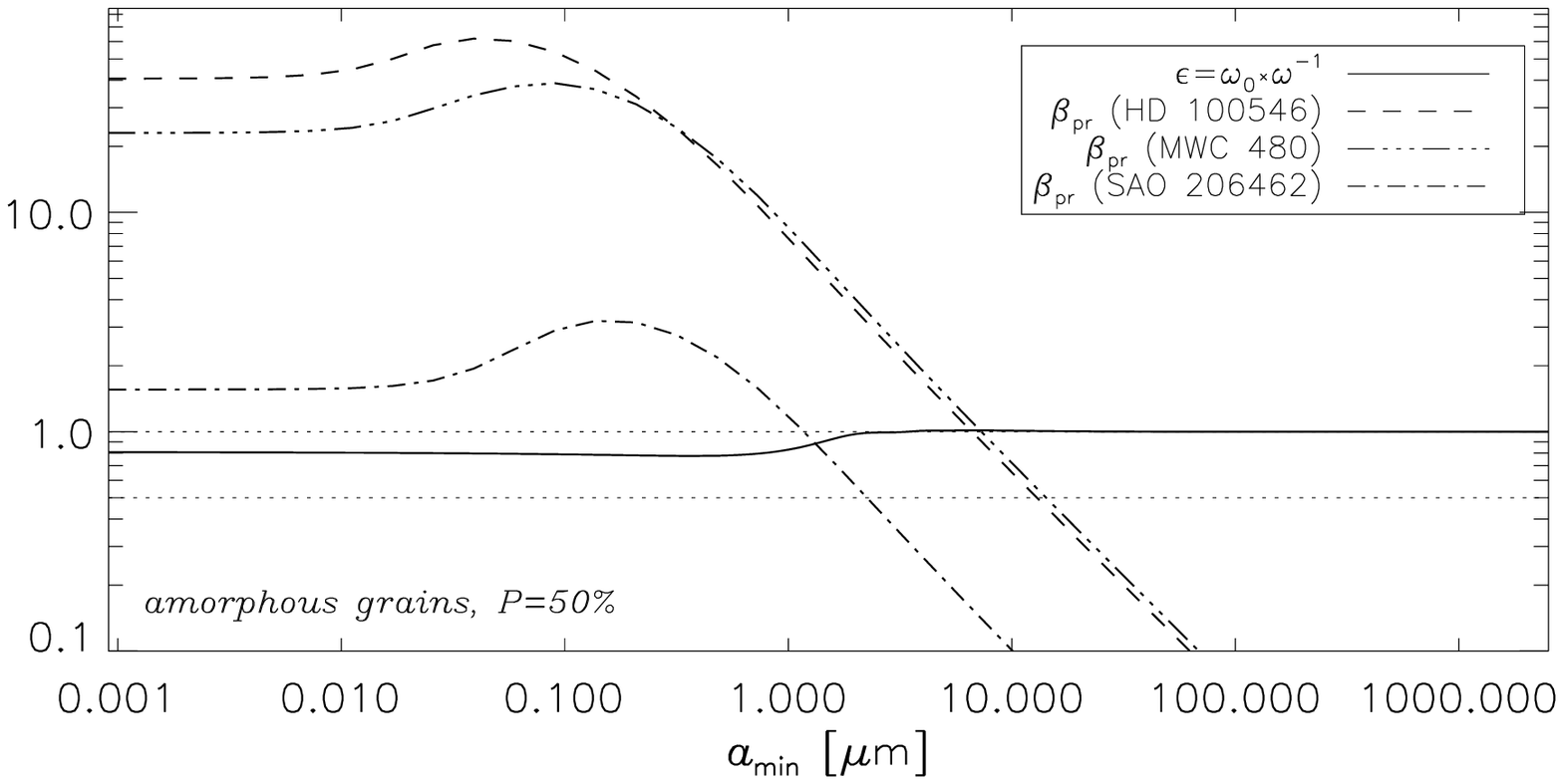}
\vspace{1truemm}\par\noindent
\includegraphics[angle=90,origin=bl,width=0.49\textwidth]{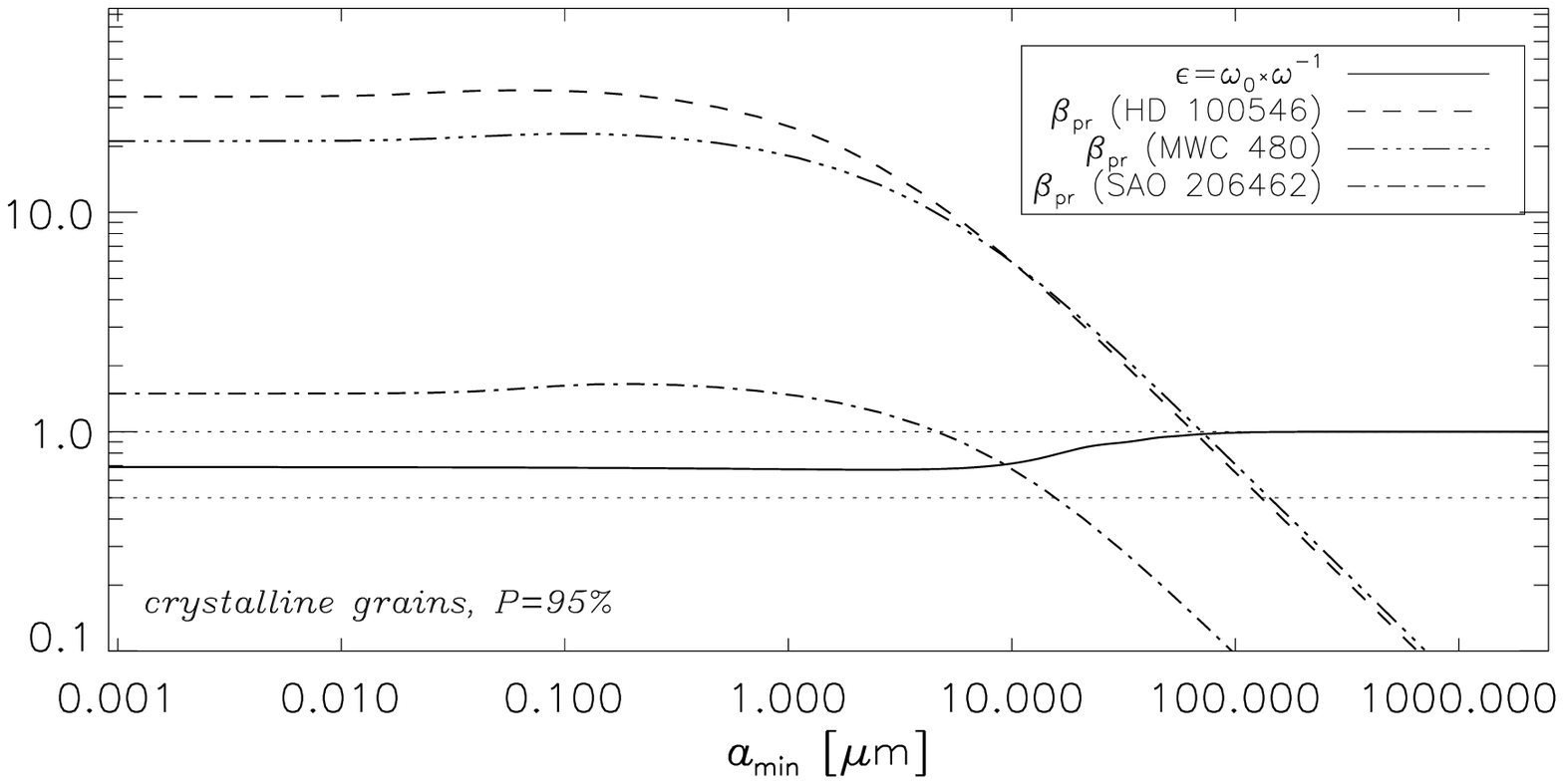}
\caption[]{Correcting factor $\epsilon$ at $\lambda=1.6\,\mu$m as a
  function of minimum grain size $a\dma{min}$ (see text). The ratios
  of radiation pressure to gravitational forces $\beta\dma{pr}$ have
  been computed for grains with size $a\dma{min}$ in the vicinity of
  \hda, \mwc\ and \hdb\ \citep[assumed to be a F4V star with
  L$_*$\,$\simeq$\,2.5\,L$_{\odot}$ and
  M$_*$\,$\simeq$\,1.2\,M$_{\odot}$,][]{all73}.}
\label{albedo}
\end{center}
\end{figure}
The $\omega$ ratio depends on the wavelength, on the grain size
distribution and on grain properties (shape, chemical composition ...
).  In the following, we assume $\lambda=1.6\,\mu$m and a collisional
grain size distribution proportional to $a^{-3.5}$ between
$a\dma{min}$ and $a\dma{max}$. Grains are assumed to be spherical ($a$
is the radius) and $a\dma{max}$ is fixed to 1\,cm. We adopt the two
types of grains as proposed in \citet{aug99a}, namely: amorphous
``ISM-like grains'' with porosity $P$ of about 50\% and crystalline
``comet-like grains'' with large porosity ($P\sim95$\%). Grains are
assumed to be made of silicates, organic refractories and a small
amount of water ice (10\%). We use the Maxwell-Garnett effective
medium theory to compute the complex index of refraction of the
aggregate and the Mie theory to derive the optical properties.

For $a\dma{min}\gg\lambda$, the $\omega$ ratio reaches a constant
value $\omega_0$ close to 0.5 \citep{boh83}. Then if we note:
\begin{eqnarray*}
\tau_{\perp , 0}(r)=  2 \, \Phi_{\lambda}^{-1}  \,
f(90\degr)^{-1} \, r^2\,SB_{\perp}(r) \\
\textrm{and}\,\,\, \epsilon = \omega_0 \times \omega^{-1}
\end{eqnarray*}
we have:
\begin{eqnarray*}
\tau_{\perp}(r) \simeq \epsilon \, \tau_{\perp , 0}(r)
\end{eqnarray*}
where $\epsilon$ represents a correcting factor which must be applied
for the smallest $a\dma{min}$, where the geometric optics approximation
does not apply. $\epsilon$ versus $a\dma{min}$ is shown
in Figure \ref{albedo} for the two types of grains. Note that for
$a\dma{min}$ larger than a few $\frac{\lambda}{2\pi (1-P)^{1/3}}$,
$\omega$ is very close to the albedo of the smallest grains.
For simplicity, we assume that grains scatter isotropically, then:
$f(90\degr) \simeq 1/4\pi$.
\section{Disk normal surface density}
The normal surface density of the disk is obtained by: 
\begin{eqnarray*}
\sigma_{\perp}(r) = \langle \frac{4}{3}\pi\rho\,a^3\rangle_a
\times n_{\perp}(r) \simeq \zeta\, \sigma_{\perp , 0}(r) \\
\textrm{with:}\,\,\, \sigma_{\perp , 0}(r) = \frac{2}{3}\rho\dma{g}
\tau_{\perp , 0}(r)\sqrt{a\dma{min}a\dma{max}} \\
\textrm{and:}\,\,\, \zeta = \frac{\langle \pi a^2\rangle_a}
{\langle \sigma\dma{sca}\rangle_a \,}\,\,\,\, ,
\end{eqnarray*}
assuming $\omega_0\simeq 0.5$. The grain density $\rho\dma{g}$ depends
on the grain properties. It is about $1.2\,$g.cm$^{-3}$ for amorphous
grains and ten times smaller for crystalline grains due to the large
porosity.

\begin{figure}[tbp]
\begin{center}
\includegraphics[angle=90,origin=bl,width=0.49\textwidth]{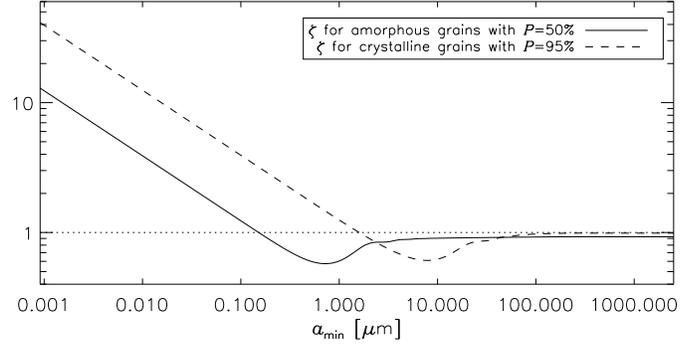}
\vspace*{0.3cm}
\caption[]{Correcting factor $\zeta$ at $\lambda=1.6\,\mu$m as a
  function of minimum grain size $a\dma{min}$ (see text).}
\label{zeta}
\end{center}
\end{figure}

The $\zeta$ ratio plotted in Figure \ref{zeta} is the correcting
factor (as $\epsilon$ for the normal optical thickness). It must be
applied for the smallest $a\dma{min}$, {\it i.e.} where the more
numerous grains (those with sizes close to $a\dma{min}$) are not
efficient scatterers. But in the same time (for the same smallest
$a\dma{min}$), $\zeta$ is proportional to $1/\sqrt{a\dma{min}}$ and
then the normal surface density does not depend any more on the
minimum grain size.

The normal surface density $\sigma_{\perp}(r)$ can therefore be
inferred from the observed surface brightness if the disk is seen
almost pole-on. If the disk is not resolved, it is also possible to
derive an upper limit on $\sigma_{\perp}(r)$ from the detection limit.
Obviously, the main uncertainty on
$\sigma_{\perp}(r)$ comes from the lack of strong constraints on the
minimum and maximum grain sizes in the disk.


\begin{thebibliography}{}
\def\aj{AJ\ }
\def\aea{A\&A\ }
\def\aeasup{A\&AS\ }
\def\apj{ApJ\ }
\def\apjl{ApJL\ }
\def\apjsup{ApJS\ }
\def\sci{Sci\ }
\def\apss{Ap\&SS\ }
%
\bibitem[Allen(1973)]{all73}Allen C.W., 1973, Astrophysical Quantities,
Third edition, eds. The Athlone Press LTD.
\bibitem[Augereau et al.(1999a)]{aug99a} Augereau J.C., Lagrange A.M.,
Mouillet D., Papaloizou J.C.B., Grorod P.A., 1999a, A\&A 348, 557
\bibitem[Augereau et al.(1999b)]{aug99b} Augereau J.C., Lagrange A.M.,
Mouillet D. \&\ M\'enard F., 1999b, A\&AL, 350, 51
\bibitem[Aumann et al.(1984)]{aum84} Aumann H.H., Gillett F.C., Beichman C.A.,
de Jong T., Houck J.R., Low F.J., Neugebauer G., Walker R.G. \&\ Wesselius P.R,
1984, ApJL 278, 23
\bibitem[Backman \&\ Paresce(1993)]{bac93} Backman D. and Paresce F.
  1993, in Protostars and Planets III, (ed. E.H. Levy and Lunine),
  Tucson: University of Arizona Press, p. 1253
\bibitem[Barrado y Navascu\'es et al.(1999)]{bar99} Barrado y Navascu\'es D.,
Stauffer J.R., Song I. \&\ Caillault J.-P., 1999, ApJL 520, 123
\bibitem[Beust et al.(2000)]{beu00} Beust H., Karmann C. \&\ Lagrange A.M.,
2000, A\&A, submitted
\bibitem[Bohren \&\ Huffman(1983)]{boh83} Bohren C.F., Huffman D.R.,
  1983, Absorption and scattering of light by small particles, Wiley,
  New-York
\bibitem[Bouwman et al.(2000)]{bou00} Bouwman J. et al., 2000, IAU
  Symposium S202, Manchester, UK
\bibitem[Cameron(1995)]{cam95} Cameron A.G.W., 1995, Meteoritics 30, 133
\bibitem[Charbonneau et al.(2000)]{cha00} Charbonneau D.B., Brown T.M.,
Latham D.W. \&\ Mayor M., 2000, ApJL 529, 45
\bibitem[Clarke et al.(1999)]{cla99} Clarke D., Smith R.A. \&\ Yudin R.V.,
1999, A\&A 347, 590
\bibitem[Coulson et al.(1995)]{cou95} Coulson I.M. \&\ Walther D.M., 1995,
MNRAS 274, 977
\bibitem[Coulson et al.(1998)]{cou98} Coulson I.M., Walther D.M. \&\
Dent W.R.F., 1998, MNRAS 296, 934
\bibitem[Crovisier et al.(1997)]{cro97} Crovisier J., Leech K.,
Bockel\'ee-Morvan D., Brooke T.Y., Hanner M.S., Altieri B., Keller H.U. \&\
Lellouch E., 1997, Science 275, 1904
\bibitem[de Winter et al.(1999)]{dew99} de Winter D., Grady C. A., van den
Ancker M. E., P\'erez M. R., Eiroa C., 1999, A\&A 343, 137
\bibitem[Dunkin et al.(1997)]{dun97} Dunkin S.K., Barlow M.J., Ryan S.G.,
1997, MNRAS 286, 604
\bibitem[Dutrey et al.(1996)]{dut96} Dutrey A., Guilloteau S., Duvert G.,
Prato L., Simon M., Schuster K. \&\ M\'enard F., 1996, A\&A 309, 493
\bibitem[Grady et al.(1996)]{gra96} Grady C.A., Perez M.R., Talavera A.,
Bjorkman K.S., de Winter D., Th\'e P.-S., Molster F.J., van den Ancker M.E.,
Sitko M.L.,Morrison N.D., Beaver M.L., McCollum B. \&\ Castelaz M.W., 1996,
A\&AS 120, 157
\bibitem[Grady et al.(1997)]{gra97} Grady C.A., Sitko M.L., Bjorkman K.S.,
Perez M.R., Lynch D.K., Russel R.W. \&\ Hanner M.S., 1997, ApJ 483, 449
\bibitem[Greaves et al.(1998)]{gre98} Greaves J.S., Holland W.S.,
Moriarty-Schieven G., Jenness T., Dent W.R.F., Zuckerman B., McCarthy C.,
Webb R.A., Butner H.M., Gear W.K. \&\ Walker H.J., 1998, ApJL 506, 133
\bibitem[Habing et al.(1999)]{hab99} Habing H. J., Dominik C., Jourdain de
Muizon M., Kessler M. F., Laureijs R. J., Leech K., Metcalfe L., Salama A.,
Siebenmorgen R., Trams N., 1999, Nature 401, 456
\bibitem[Heap et al.(2000)]{hea00} Heap S.R., Lindler D.J., Lanz T.M.,
Cornett R.H., Hubeny I., Maran S.P., Woodgate B., 2000, ApJ 530
\bibitem[Henning et al.(1994)]{hen94} Henning Th., Launhardt R.,
  Steinacker J. \&\ Thamm E., 1994
\bibitem[Henning et al.(1998)]{hen98} Henning Th., Burkert A., Launhardt R.,
  Leinert C. \&\ Stecklum B., 1998, A\&A 336, 565
\bibitem[Holland et al.(1998)]{hol98} Holland W.S., Greaves J.S.,
Zuckerman B., Webb R.A., McCarthy C., Coulson I.M., Walther D.M., Dent W.R.F.,
Gear W.K. \&\ Robson I., 1998, Nature 392, 788
\bibitem[Hu et al.(1989)]{hu89} Hu J.Y., Th\'e P.S. \&\ de Winter D.,
1989, A\&A 208, 213
\bibitem[Jayawardhana et al.(2000)]{jay00} Jayawardhana R., Fisher R.S.,
Telesco C.M., Pi\~na R.K., Barrado y Navascues D., Hartmann L., Fazio G.,
2000, PhD thesis
\bibitem[Kenyon \&\ Hartmann(1997)]{ken97} Kenyon S.J.\&\ Hartmann L., 1997,
ApJ 323, 714
\bibitem[Krist et al.(2000)]{kri00} Krist J.E., Stapelfeldt K.,
  M\'enard F., Padgett D., Burrows C., 2000, ApJ 538, 793
\bibitem[Lagrange et al.(2000)]{lag00} Lagrange A.M., Backman D.,
Artymowicz P., 2000 in PPIV, pp 639-672
\bibitem[Lecavelier des Etangs et al.(1996)]{lec96} Lecavelier des Etangs A.,
Vidal-Madjar A. \& Ferlet R., 1996, A\&A 307, 542 
\bibitem[Lin \&\ Pringle(1990)]{lin90} Lin D.N.C., \&\ Pringle J.E., 1990,
ApJ 358, 515
\bibitem[Malfait et al.(1998)]{mal98} Malfait K., Waelkens C., Waters
L.B.F.M., Vandenbussche B., Huygen E. \&\ de Graauw M.S., 1998, A\&AL 332, 25
\bibitem[Mannings \&\ Sargent(1997a)]{man97a} Mannings V., \&\ Sargent A.I.,
1997a, ApJ 490, 792
\bibitem[Mannings et al.(1997b)]{man97b} Mannings V., Koerner D.W.,
Sargent A.I., 1997b, Nature 388, 555
\bibitem[Miroshnichenko et al.(1997)]{mir97} Miroshnichenko A., Ivezi\'c Z.
\&\ Elitzur M., 1997, ApJL 475, 41
\bibitem[Miroshnichenko et al.(1999)]{mir99} Miroshnichenko A., Ivezi\'c Z.,
Vinkovi\'c D. \&\ Elitzur M., 1999, ApJL 520, 115
\bibitem[Myers et al.(1998)]{mye98} Myers J.R., Sande C.B., Miller A.C.,
Warren Jr. W.H., Tracewell D.A., 1998, Sky2000 Master Catalog, Goddard Space
Flight Center, Flight Dynamics Division
\bibitem[Nakano(1990)]{nak90} Nakano T., 1990, ApJL 355, 43 
\bibitem[Padgett et al.(1999)]{pad99} Padgett D.L., Brandner W., Stapelfeldt
K.R., Strom S.E., Terebey S. \&\ Koerner D., 1999, AJ 117, 1490-1504
\bibitem[Pantin et al.(2000)]{pan00} Pantin E., Waelkens C., Lagage
P.O., 2000, A\&AL, in press
\bibitem[Queloz et al.(2000)]{que00}Queloz D., Mayor M., Weber L., Blecha A.,
Burnet M., Confino B., Naef D., Pepe F., Santos N.C., Udry S., 2000,
A\&A 354, 99
\bibitem[Schultz et al.(1999)]{sch99} Schultz A.B., Storrs A.D.,
  Fraquelli D., Instrument Science Report NICMOS-99-006
\bibitem[Sitko et al.(1999)]{sit99} Sitko M.L., Grady C.A., Lynch D.K.,
Russell R.W., Hanner M.S., 1999, ApJ 510, 408
\bibitem[Sylvester et al.(1996)]{syl96} Sylvester R.J., Skinner
C.J., Barlow M.J. \&\ Mannings V., 1996, MNRAS 279, 915
\bibitem[Sylvester et al.(1997)]{syl97} Sylvester R.J., Skinner
C.J. \&\ Barlow M.J., 1997, MNRAS 289, 831
\bibitem[van den Ancker et al.(1997)]{van97} van den Ancker M.E., Th\'e P.S.,
Tjin A Djie H.R.E., Catala C., de Winter D., Blondel P.F.C. \&\ Waters L.B.,
1997, A\&A 324, L33
\bibitem[van den Ancker et al.(1998)]{van98} van den Ancker M.E., de
Winter D., Tjin A Djie H.R.E., 1998, A\&A 330, 145
\bibitem[Vieira et al.(1999)]{vie99} Vieira S.L.A., Pogodin M.A. \&\
Franco G.A.P., 1999,  A\&A 345, 559
\bibitem[Waelkens et al.(1990a)]{wae90a} Waelkens C., Engelsman E., Waters L.B.F.M., Van der Veen W.E.C.J. \&\ Trams N.R., 1990a, in "From Miras to Planetary Nebulae: Wich Path for Stellar Evolution?", eds. Menessier M.O. and Omont A., Editions Fronti\`eres, France, p.470
\bibitem[Waelkens et al.(1990b)]{wae90b} Waelkens C., Van Winckel H. \&\ Trams N.R., 1990b, in IAU Symp. 145, "Evolution of stars: The Photospheric Abundance Connection", Poster Papers, p.21, eds. Michaud, Tutukov \&\ Bergevin, Montreal, Canada
\bibitem[Waters \&\ Waelkens(1998)]{wat98} Waters L.B. \&\ Waelkens C.,
1998, ARA\&A 36, 233
\bibitem[Weinberger et al.(1999)]{wei99} Weinberger A.J., Becklin E.E.,
Schneider G., Smith B.A., Lowrance P.J., Silverstone M.D., Zuckerman B.,
Terrile R. J., 1999, ApJL 525, 53
\bibitem[Yudin \&\ Evans(1998)]{yud98} Yudin R.V. \&\ Evans A.,
1998, A\&AS 131, 401
\bibitem[Zuckerman \&\ Becklin(1993)]{zuc93} Zuckerman B. \&\ Becklin E.E.,
1993, ApJ, 414, 793
\bibitem[Zuckerman et al.(1995)]{zuc95} Zuckerman B., Forveille T. \&\
Kastner J.H., 1995, Nature 373, 494
%
\end{thebibliography}
\end{document}